\documentclass[12pt,tightenlines,eqsecnum,floats,aps,amsmath,amssymb,nofootinbib,prd,shownopacs,floatfix]{revtex4}

\usepackage{setspace}
\usepackage{amssymb}
\usepackage{graphicx,wrapfig}
\usepackage{enumerate} 
\usepackage{color}
\usepackage{mathrsfs}
\usepackage{subfigure}
\definecolor{mg}{rgb}{0.0, 0.5, 0.0}

\def\be{\nopagebreak[3]\begin{equation}}
\def\ee{\end{equation}}
\def\ba{\nopagebreak[3]\begin{eqnarray}}
\def\ea{\end{eqnarray}}
\newcommand{\f}{\frac}

\def\lp{\ell_{\rm Pl}}
\def\d{{\rm d}}

\def\fq{\mathring{q}_{ab}}
\def\t{\tilde}

\def\db{\delta_b}
\def\dc{\delta_c}
\def\T{\mathcal{T}}
\def\GammaE{\Gamma_{\rm ext}}
\def\GammaEb{\bar\Gamma_{\rm ext}}
\def\GammaEh{\hat\Gamma_{\rm ext}}
\def\Hee{H_{\rm eff}^{\rm ext}}
\def\H{\mathcal{H}}

\begin{document}

\title{Quantum Extension of the Kruskal Space-time}

\author{Abhay Ashtekar$^{1}$, Javier Olmedo$^{1}$, Parampreet Singh$^{2}$}
\affiliation {
1. Institute for Gravitation and the Cosmos, Penn State University,
University Park, PA 16801 \\
2. Department of Physics and Astronomy, Louisiana State University,
Baton Rouge, LA 70803}

\begin{abstract}
A new description of macroscopic Kruskal black holes that  incorporates   the quantum geometry corrections of loop quantum gravity is presented. It encompasses both the `interior' region that contains classical singularities and the `exterior' asymptotic region.  Singularities  are naturally resolved by the quantum geometry effects of loop quantum gravity. The resulting quantum extension of space-time has the following features: (i) It admits an infinite number of trapped, anti-trapped and asymptotic regions; (ii) All curvature scalars have  uniform (i.e., mass independent) upper bounds; (iii) In the large mass limit, all asymptotic regions of the extension have the same ADM mass; (iv) In the low curvature region (e.g., near horizons) quantum effects are negligible, as one would physically expect; and (v) Final results are insensitive to the fiducial structures that have to be introduced to construct the classical phase space description (as they must be).  Previous effective theories \cite{ab,lm,bv,dc,cgp,bkd,cs,djs,oss,cctr} shared some but not all of these features. We compare and contrast our results with those of these effective theories and also with expectations based on the  AdS/CFT conjecture \cite{eh}.  We conclude with a discussion of limitations of our framework, especially for the analysis of evaporating black holes.

\end{abstract}
\maketitle

\section{Introduction}
\label{s1}
It is widely believed that predictions of general relativity cannot be trusted once space-time curvature enters the Planck regime since modifications to Einstein's equations due to quantum gravity effects would then begin to dominate. In particular, singularities of classical general relativity are often regarded as windows onto new physics. In loop quantum gravity (LQG),  new physics emerges from the underlying  \emph{quantum} Riemannian geometry  (see, e.g., \cite{apbook})  Thus, for example, in the commonly used cosmological models singularities are naturally resolved because, once a curvature invariant approaches the Planck scale, quantum geometry modifications of Einstein dynamics introduce strong `repulsive corrections' that dilute that invariant, preventing a blow-up \cite{asrev,ps}.

It is then natural to ask  if these quantum geometry effects also resolve black hole singularities. The simplest model is provided by the Schwarzschild-Kruskal space-time. For the question of singularity resolution, it suffices to restrict oneself to the black hole region that is bounded by  the singularity and event horizons, often referred to as the \emph{Schwarzschild interior}.  Since this region is isometric to the (vacuum) Kantowski-Sachs cosmological model, one can transport LQG techniques developed for homogeneous but anisotropic cosmologies. Therefore the Schwarzschild interior has drawn considerable attention from the LQG community (see, e.g.   \cite{ab,lm,bv,dc,ck,cgp,bck,sk,bkd,djs,cs,oss,yks,cctr}  for investigations closely related to this paper). The general procedure used in all these investigations is the same: (A) The classical theory is cast in a Hamiltonian framework using connection variables; (B) The passage to quantum theory is through a representation of the fundamental commutation relations that `descends' from  full LQG, and  therefore has in-built elements of quantum geometry; (C)  The quantum Hamiltonian constraint  is constructed by replacing curvature with holonomies of the gravitational connection around suitable loops that enclose minimum non-zero area allowed by  quantum geometry; and, finally,  (D) Detailed physical predictions of the model are obtained using certain `effective equations.%
\footnote{\label{fn1} In the cosmological models, effective equations were first introduced by examining the form of the quantum Hamiltonian constraint, then writing down an effective Hamiltonian constraint on the \emph{classical} phase space that  includes key quantum corrections due to quantum geometry effects of LQG, and calculating its dynamical flow, again on the classical phase space. Later these equations were shown to follow from the full quantum dynamics of sharply peaked states \cite{jw,vt,asrev}.  In the Schwarzschild case this last step has not been carried out in any of the approaches, including ours.}
Solutions to these equations show that the central singularity is resolved due to quantum corrections. We follow the same procedure. As we show in section \ref{s3}, the singularity is replaced by a space-like, 3-dimensional  \emph{transition surface} $\T$ to the past of which we have a trapped region (as in the Schwarzschild-Kruskal black hole  region) and to the future of which is an anti-trapped region (as in the Schwarzschild-Kruskal white hole  region).

However the analyses  \cite{ab,lm,bv,dc,ck,cgp,bck,sk,bkd,djs,cs,oss,yks,cctr}  differ in the way  step (C) is implemented in detail. Consequently, the resulting effective dynamics of step (D) varies from one approach to another.   Subsequent investigations have revealed that these effective descriptions have undesirable or puzzling features whose physical origin has remained unclear. For example,  physical effects fail to be independent of the fiducial structure introduced to construct the classical phase space in some approaches \cite{ab,lm,cgp}, while  quantum geometry effects could be large in low curvature regimes in other approaches. In particular,  the quintessentially quantum  transition surface could emerge in a low curvature region for macroscopic black holes with large masses \cite{cs,oss}.  Similarly, space-time geometry near the black hole horizon could receive large  quantum corrections even when the mass of the black hole is very large and hence the curvature near its horizon is low \cite{bv,dc,cctr}.%
 \footnote{ The same turns out to be true for the $\db,\dc$  prescriptions studied in  Refs. \cite{dc2,js} in the context of Kantowski-Sachs cosmologies, when they are used to model the Schwarzschild interior.}  
 A systematic discussion of these limitations, including their origin,  is given in sections \ref{s4.D}  (and \ref{s6}).

To compare and contrast these investigations, it is convenient to divide them into three broad classes,  in terms of their method of selecting the loops needed in step (C).  In all these treatments,  the loops are characterized by two quantum parameters, labeled $\delta_{b}$ and $\delta_{c}$.   In \cite{ab,lm,cgp} these parameters are set to a constant; in \cite{cs,oss} they are certain Dirac observables, i.e.,  functions on phase space that are constant along (effective) dynamical trajectories; and  in \cite{bv,dc,dc2,bkd,js,djs,cctr}  they are more general functions on the classical phase space that change not only from one dynamical trajectory to another but also along each individual trajectory. The overall strategy we will adopt is the same as that in \cite{cs} but the specific Dirac observables we use are chosen more judiciously, using conditions that refer to the transition surface $\T$. As a result,  unlike in \cite{cs}, the transition surface $\T$ always lies in the Planck regime in our effective theory, and there is also excellent agreement with classical general relativity in low curvature regions.  The trapped and anti-trapped regions are joined smoothly to asymptotic regions, leading to a genuine quantum extension of the full Kruskal space-time beyond classical singularities. (For a Penrose diagram of the full extension, see Fig. \ref{fig:4}).  

There is another  key difference from previous investigations. The primary focus  there was on Kantowski-Sachs space-times, with emphasis on issues that feature prominently in anisotropic \emph{cosmologies}, such as bounces of various scale factors {\cite{bv,dc2,bkd,cs,djs,cctr}, behavior of the energy density, expansion scalar,  shear potentials of the Weyl curvature {\cite{js}}, and, geodesic completeness and generic resolution of strong singularities \cite{ss}.}  Some of the discussions also included matter sources {\cite{dc2,js,ss}} or a cosmological constant {\cite{bkd,djs}}. While { the inclusion of matter} is natural from the cosmological perspective, the analysis no longer has direct relevance for quantum modifications of the Schwarzschild geometry near its singularity. Finally,  {a limitation of all existing studies of loop quantization of the Schwarzschild interior is that effective geometry is not extended to the asymptotic regions.} By contrast, the object of primary interest to this investigation is the Kruskal space-time, and the emphasis is on the corresponding  space-time notions, such as trapped and anti-trapped regions, black hole type and white hole type horizons, the corresponding asymptotic regions, and the behavior of the static Killing field as one passes from the original Kruskal space-time to its quantum extension. In particular, we also introduce an effective description in the asymptotic regions and show that the effective metric in the asymptotic and interior regions match at the horizons.

Material is organized as follows. Section \ref{s2} fixes the notation used subsequently, recalls the basics of the Hamiltonian framework, and summarizes the effective equations and their solutions.  This discussion is included to make the paper self-contained. In section \ref{s3} we  shift the focus from phase space trajectories to space-time geometry and discuss the causal structure of the effective space-time metric. As one might expect, corrections to the classical geometry are large in the regime where space-time curvature enters Planck scale. As a result, a transition surface $\T$ with regular space-time geometry now emerges, separating trapped and anti-trapped regions. There is a precise sense in which  $\T$ replaces the classical singularity in the quantum corrected, effective geometry. In section \ref{s4} we motivate and introduce our specific choice of the quantum parameters $\delta_{b}, \delta_{c}$ and discuss various features of the resulting effective space-time geometry in the extended Schwarzschild interior.  In section \ref{s5}  we introduce the effective description in the exterior, asymptotic region of Schwarzschild space-time and show that this effective metric matches smoothly to that in the interior.%
\footnote{Another approach that incorporates both the `interior' and the `exterior' regions of Schwarzschild but without making a direct use of homogeneity is summarized in \cite{gop}. This approach has also been used to study effects of quantum geometry on Hawking radiation \cite{gp-hawking}, and collapse of self-gravitating shells \cite{cgop}.}
For macroscopic black holes space-time curvature is small near horizons.  As one would hope on physical grounds,  quantum corrections are also small near horizons  in both interior and exterior regions and further decay rapidly as one moves away from the horizon in the asymptotic region. In section \ref{s6}  we summarize our main results and contrast our approach and findings with those used in previous investigations in LQG and also with the expectations based on the AdS/CFT correspondence \cite{eh}.  We conclude with a discussion of limitations of our analysis.

Since the LQG literature on quantum corrections to the Schwarzschild metric  is spread over 10-15 years,  we have made an attempt to make the paper self contained by recalling key ideas at various junctures. Throughout, one should bear in mind that effective descriptions can be expected to provide a good approximation to the quantum evolution only if the mass $M$ of the black hole is large, i.e., $GM =: m \gg \lp$, where $\lp$ is the Planck length. This is the regime of interest to this paper.

\section{Preliminaries}
\label{s2} 

As explained in Sec. \ref{s1}, in the first part of the paper we will focus on the Schwarzschild interior and use the fact that this portion of the Kruskal space-time is naturally foliated by a family of homogeneous space-like 3-manifolds. One can therefore use techniques from loop quantum cosmology (LQC) to construct a Hamiltonian framework based on connection and triad variables, $A^i_a$ and $E^a_i$, and then pass to the quantum theory using the same methods that are used in full LQG  (see, e.g., \cite{asrev}).  In cosmological models quantum dynamics leads to a non-singular evolution in which matter density and curvature remain finite. Furthermore, one can extract an effective description \cite{jw,vt,asrev} from the resulting quantum theory following a systematic procedure based on a geometrical formulation of quantum mechanics \cite{gqm}. Rigorous numerical simulations have shown that the effective description provides an excellent approximation to the underlying quantum dynamics for isotropic \cite{aps3,numlsu-2,numlsu-3}, as well as anisotropic space-times \cite{numlsu-4}   so long as we consider quantum states that are sharply peaked  on  `large' universes at late times.

For the Schwarzschild interior, one can also introduce quantum kinematics (see e.g., \cite{ab,cs}) and write down the Hamiltonian constraint operator. However,  for our choice of quantum parameters $\delta_{b}, \, \delta_{c}$, its explicit action is rather complicated (see Appendix \ref{a1}). Therefore, in this paper \emph{we will focus only on the effective description}, leaving the exploration of  its relation to full quantum dynamics for a future work. In contrast to previous investigations of effective dynamics, our emphasis will be more on the causal structure and geometric properties of the quantum corrected space-time than on phase space trajectories and cosmological issues associated with the Kantowski-Sachs spacetime. The purpose of this section is to summarize this procedure. We begin with the phase space and classical dynamics and then discuss effective dynamics. Our treatment and conventions are based on Refs. \cite{ab,cs}, which the reader can refer to for further details.

\subsection{Phase space and classical dynamics}
\label{s2.A}

Recall that the interior of the Kruskal space-time is isometric to the  Kantowski-Sachs vacuum solution. The homogeneous Cauchy slices have topology $\mathbb{R} \times \mathbb{S}^2$. As is customary in the phase space formulation of homogeneous models, let us  introduce a fiducial metric $\fq$ on $\Sigma$ 
\be \label{fiducial}
\fq \d x^a \d x^b = \d x^2 + r_o^2(\d \theta^2 + \sin^2 \d \phi^2) ,
\ee
where $x \in (-\infty,\infty)$,  $\theta$ and $\phi$ are 2-sphere coordinates, and $r_{o}$ is a constant (with dimensions of length).  Since $\Sigma$ is non-compact in the $x$ direction, and all fields under consideration are homogeneous, in the construction of the phase space description we need to introduce an infrared cut-off; otherwise expressions of the symplectic structure and the (integrated) Hamiltonian constraint would be divergent. This is achieved by introducing a fiducial cell $\mathcal{C}$ in $\Sigma$, also with topology $\mathbb{R}\times\mathbb{S}^{2}$, but with $x \in (0, L_{o})$.  In the phase space considerations all fields and integrals will be restricted to $\mathcal{C}$. Although intermediate structures will refer to $L_{o}$, the final physical results --such as equations of motion-- will be independent of this choice in our classical as well as effective theory.

Using the underlying spatial homogeneity of the Kantowski-Sachs space-time, we can solve the spatial diffeomorphism constraint and perform a partial gauge fixing to satisfy the Gauss constraint. As a result,  the gravitational connection and the conjugate densitized triad can be expressed as
\be 
A^i_a \, \tau_i \, \d x^a \, = \, \bar c \, \tau_3 \, \d x + \bar b\,r_o \, \tau_2 \d \theta 
- \bar b \,r_o\, \tau_1 \sin \theta \, \d \phi + \tau_3 \cos \theta \, \d \phi,
\ee
and
\be 
E^a_i \, \tau^i \partial_a \, =  \, \bar p_c \,r_o^2\, \tau_3 \, \sin \theta 
\, \partial_x + \bar p_b\, r_o \, \tau_2 \, \sin \theta \, 
\partial_\theta - \bar p_b\, r_o \, \tau_1 \,  \partial_\phi . 
\ee
Here $\tau_i$ are SU(2) generators related to Pauli spin matrices $\sigma_{i}$ via  $\tau_i = - i \sigma_i/2$, and  the constants $\bar{b},\bar{c}, \bar{p}_{b}, \bar{p}_{c}$ represent the dynamical variables. Thus, the symmetry reduced phase space is now coordinatized by two configuration variables $(\bar{b}, \, \bar{c})$ and their conjugate momenta $(\bar p_b, \bar p_c)$. The symplectic structure is given by
\be
\bar \Omega = \frac{L_o r_o^2}{2 G \gamma} \left(2 \d \bar b \wedge \d \bar p_b + \d \bar c \wedge \d \bar p_c \right),
\ee
where  $\gamma$ is the Barbero-Immirzi parameter  that captures the quantization ambiguity in 
LQG.%
\footnote{\label{fn4} The parameter  $\gamma$ arises in the passage from classical to quantum theory.  In quantum theory, it  determines the LQG area gap $\Delta$ via $\Delta = 4\sqrt{3} \pi\, \gamma\, \lp^{2}$. Its value is generally fixed to be $\gamma = 0.2375$ using black hole entropy considerations. Although in the final picture one can get rid of $\gamma$ in favor of the more fundamental and physical parameter $\Delta$,  both parameters feature 
in various expressions in the existing literature \cite{ab,lm,dc,bv,cs}.  To facilitate comparison we will also keep both parameters.}
 The symplectic structure depends explicitly on the length $L_o$ of the fiducial cell and the radius $r_o$ in the fiducial metric $\fq$. We can absorb  $L_o$ and $r_o$ by rescaling  the connection and triad variables: $b = r_o \bar b$, $c = L_o \bar c$, $p_b = L_o r_o \bar p_b$ and 
$p_c = r_o^2 \bar p_c$. They then satisfy the following Poisson brackets:
\be\label{pbs}
\{c,\,p_c\} \, = \,2 G \gamma, \quad \{b,\,p_b\} \, = \,  G \gamma.
\ee
Note that under the transformation $r_o \rightarrow \beta r_o$  (where $\beta$ is a constant)  the rescaled connection and triad variables are invariant. However, under a rescaling of fiducial length $L_o$: $L_o \rightarrow \alpha L_o$ (where $\alpha$ is a constant), we get $b \rightarrow b$, $c\rightarrow \alpha c$, $p_b \rightarrow \alpha p_b$ and $p_c \rightarrow p_c$. Therefore, physical quantities can depend only on $b, p_{c}$ and the combinations $c/L_{o}$ and $p_{b}/L_{o}$.

The gravitational connection and spatial triads now take the form:
\be \label{connection}
A^i_a \, \tau_i \, \d x^a \, = \, \f{c}{L_{o}}  \, \tau_3 \, \d x +  b \, \tau_2 \d \theta 
- b\, \tau_1 \sin \theta \, \d \phi + \tau_3 \cos \theta \, \d \phi,
\ee
and
\be \label{triad}
E^a_i \, \tau^i \partial_a \, =  \,  p_c \, \tau_3 \, \sin \theta 
\, \partial_x + \f{p_b}{ L_o} \, \tau_2 \, \sin \theta \, 
\partial_\theta -\f {p_b}{ L_o} \, \tau_1 \,  \partial_\phi . 
\ee
Given any choice of a time coordinate $\tau$ and the associated lapse $N_{\tau}$, 
each point in the phase space defines a spatially homogeneous metric with Kantowski-Sachs isometries via
\be\label{metric}
g_{ab} \d x^{a} \d x^{b} \equiv \d s^2 = - N_{\tau}^2 \d \tau^2 + \f{p_b^2}{|p_c| L_o^2} \d x^2 + |p_c| (\d \theta^2 + \sin^2\theta \d \phi^2) ,
\ee
Therefore, by restricting ourselves to $\tau <2m$, we can use the standard form of the \emph{interior} Schwarzschild solution 
\be\label{metric-kruskal}
\d s^2 = - \Big( \frac{2 m}{\tau} - 1 \Big)^{-1} \d \tau^2 + \Big(\frac{2 m}{\tau} - 1\Big) \d x^2 + \tau^2 (\d \theta^2 + \sin^2\theta \d \phi^2),
 \ee
(where  $m=GM$) to set up a dictionary between phase space variables and their space-time counterparts.%
\footnote{  \label{fn5} We have used the notation that is tailored to the Schwarzschild interior. The standard Schwarzschild form is obtained by substitutions $\tau \to r$  and $x \to t$.}
In terms of  $m$ and  the radius $\tau$ of metric 2-spheres we have:  $|p_c| = \tau^2$, $p_b^{2} =  L_o^{2} (\tfrac{2m}{\tau} -1)  \tau^{2}$ and   $N_{\tau}^{2} = (\f{2m}{\tau} -1)^{-1}$.

With the Gauss and spatial diffeomorphism constraint fixed, we can extract classical dynamics from the Hamiltonian constraint using Hamilton's equations.  It turns out that in the effective theory  there is a particularly convenient choice of lapse for which dynamical equations simplify sufficiently to obtain explicit solutions \cite{cs}. The classical analog of that lapse is
\be \label{Ncl}  N_{\rm{cl}} = \gamma \, b^{-1}\mathrm{sgn}(p_c) \, |p_c|^{1/2}. \ee
and we will denote the corresponding time variable by $T_{\rm cl}$. The corresponding Hamiltonian constraint is 
\be\label{Hcl}
H_{\mathrm{cl}}[N_{\rm{cl}}] = - \f{1}{2 G \gamma}\left(2 c \, p_c + \left(b + \f{\gamma^2}{b}\right) p_b  \right) .
\ee
Using Hamilton's equations, the evolution equations for connection variables turn out to be
\be \label{conf-dot}
\dot b  = G \gamma \frac{\partial H_{\mathrm{cl}}[N_{\rm{cl}}]}{\partial p_b} = -\frac{1}{2 b}\, (b^2 + \gamma^2) , \quad\mathrm{and} \quad \dot c   = 2 G \gamma \frac{\partial H_{\mathrm{cl}}[N_{\rm{cl}}]}{\partial p_c} = - 2 c~,
\ee
where the `dot' denotes time derivative with respect to $T_{\rm{cl}}$.  Similarly, for the triad variables we obtain,
\be \label{mom-dot}
\dot p_b = - G \gamma \frac{\partial H_{\mathrm{cl}}[N_{\rm{cl}}]}{\partial b} = \frac{p_b}{2 b^2}\,(b^2-\gamma^2), ~~~ \mathrm{and} ~~~ \dot p_c  =  -2 G \gamma \frac{\partial H_{\mathrm{cl}}[N_{\rm{cl}}]}{\partial c} = 2 p_c~.
\ee
Solutions to these dynamical equations together with the Hamiltonian constraint turn out to be
\be \label{conf} 
b(T_{\rm{cl}})=\pm \gamma\, \left(e^{-T_{\rm{cl}}}-1\right)^{1/2}, ~~~c(T_{\rm{cl}}) =   \mp\,c_o\,e^{-2 T_{\rm{cl}}},
\ee
and
\be \label{momenta}
p_b(T_{\rm{cl}})=p_b^{(o)}\, e^{T_{\rm{cl}}}\,\big(e^{-T_{\rm{cl}}} - 1\big)^{1/2}, \qquad p_c(T_{\rm{cl}}) = p_c^{(o)}\,e^{2 T_{\rm{cl}}} ~.
\ee
In writing these solutions, we have fixed one of the integration constants  so that, in the space-time picture, the black hole horizon lies at $T_{\rm cl}=0$.  The singularity now occurs at $T_{\rm cl} = -\infty$ so that the Schwarzschild interior corresponds to $-\infty <T_{\rm cl} <0$. The remaining three integration constants, $c_{o}, p_{b}^{(o)}$ and $p_{c}^{(o)}$ are subject to one condition coming from the Hamiltonian constraint $H_{\mathrm{cl}}[N_{\rm{cl}}] =0$ and can therefore be parametrized by two constants, $m, L_{0}$ \cite{ab,cs} : $c_{o}= \gamma L_{o}/4m; \,\, p_{b}^{(o)} = - {2m} L_{0};\,\, p_{c}^{(o)} = 4m^{2}$. Here, and in what follows we fix the orientation of the spatial triad  (see (\ref{triad}))  and restrict ourselves to $p_{c} \ge 0, c >0, b>0$ and  $p_{b} \le 0$ .

The form of the solutions immediately implies that  $c p_c/(L_o \gamma)$ is a Dirac observable --i.e. a constant of motion-- that equals $m=GM$ in the space-time metric defined by the dynamical trajectory: 
\be  \frac{cp_{c}}{L_{o}\gamma} = m \qquad {\hbox{\rm along any classical dynamical trajectory}}~. \ee
Finally, to display the standard form (\ref{metric-kruskal}) of the metric, it suffices to change the time coordinate and set $\tau := 2 m e^{T_{\rm{cl}}}$. At the horizon, identified by $\tau = 2m$, or $T_{\rm cl}=0$, we have  $b=0$ and $p_b=0$  and $c$ and $p_c$ take the values, $c(0) = \gamma L_o/4m$ and $p_c(0) = 4 m^2$. The central singularity occurs at $\tau=0$ or $T_{\rm{cl}} \rightarrow -\infty$. Here the connection components diverge and both of the triad components vanish. 

\subsection{Effective dynamics}
\label{s2.B}

Effective equations are formulated on the same phase space as the one used in the classical theory  but they incorporate leading order quantum corrections through `quantum parameters' $\delta_{b}, \delta_{c}$.  As mentioned in section \ref{s1}, we will assume  that: (1)  $\delta_b$ and $\delta_c$ are judiciously chosen Dirac observables (and thus commute with the Hamiltonian constraint); and, (2)  go to zero in the limit in which the area gap $\Delta$ is sent to zero.  Condition (1) is a subtle requirement because $H_{\rm eff}(N)$ itself depends on $\delta_{b},\delta_{c}$ (see Eq. (\ref{H_eff})). However, as our discussion in Appendix \ref{a1}  (and section \ref{s4.A})  shows, a large family of consistent choices does exist. Thus, $\delta_{b}, \delta_{c}$ will be  $\hbar$-dependent phase space functions which remain constant along dynamical trajectories.  A specific choice will be made in section \ref{s3} and it  will ensure $\delta_b\ll 1$ and $\delta_c\ll 1$ for macroscopic black holes.

Now, as mentioned in section \ref{s2.A},  a convenient choice of lapse considerably simplifies the analysis of effective dynamics and enables one to write solutions in a closed analytic form.  Therefore, following \cite{cs}, we will set%
\footnote{\label{fn6}  In the quantum theory, only holonomies defined by the connection are well-defined; not the connections themselves. As in LQC, holonomies are almost periodic functions of connections and we are led to restrict the phase space to the sector $\delta_{b} b\in (0, \pi); \, \delta_{c} c\in (0, \pi);\,\, p_{b} <0$ and $p_{c} >0$, where the last two conditions are the same as in the classical theory.}
\be \label{N}  N = \frac{\gamma \,p_c^{1/2} \,\,\delta_b}{\sin(\db b)},\ee
and we will denote the corresponding time variable by $T$.  (Just as $T_{\rm cl} <0$ in the Schwarzschild `interior', we will see that $T<0$ in the `extended Schwarzschild interior' because as in the classical theory $N$ blows up at $T=0$.) The resulting effective Hamiltonian is given by
\be \label{H_eff} 
H_{\mathrm{eff}}[N] =  - \f{1}{2 G \gamma} \Big[2 \f{\sin (\delta_c c)}{\delta_c}  \, p_c  + \Big(\f{\sin (\delta_b b)}{\delta_b} + \frac{\gamma^2 \delta_b}{\sin(\delta_b b)} \Big) \, p_b  \Big] .  \ee
It is easy to see that in the classical limit $\delta_b \rightarrow 0$ and $\delta_c \rightarrow 0$, the lapse $N$ in the effective theory  agrees with the lapse $N_{\rm{cl}}$ in the classical theory (see Eqs. (\ref{N}) and (\ref{Ncl})), and the effective Hamiltonian $H_{\mathrm{eff}}[N]$ reduces to the classical Hamiltonian (\ref{Hcl}). As shown in Appendix \ref{a1},  there exists  a class of quantum parameters $\delta_{b}, \delta_{c}$ which lead to the  the following dynamical equations for connection and triad components:
\be \label{eom1}
\dot b = - \f{1}{2} \left(\f{\sin(\db b)}{\db} +\f{\gamma^2  \db}{\sin(\db b)}\right) , ~~~~ \dot c = - 2 \, \f{\sin(\dc c)}{\dc},
\ee
and
\be \label{eom2}
\dot p_b = \f{p_{b}}{2} \, \cos(\db b) \left(1 - \f{\gamma^2  \db^2}{\sin^2(\db b)}\right) , ~~~~ 
 \dot p_c = 2 \, p_c \, \cos(\dc c) .
\ee
where the `dot' denotes derivative  with respect to $T$.

An interesting feature of the above set of equations is that  dynamics of  $b$ and $p_b$ decouples from that of $c$ and $p_c$. Thus, the trajectories for the  $(b,p_b)$ sector in the effective phase space can be obtained independently from the  trajectories for the $(c,p_c)$ sector. This feature, which is shared with the classical theory, is tied to $\db$ and $\dc$ being Dirac observables and plays a  crucial role in obtaining closed form solutions in the effective theory.  If $\db$ and $\dc$ had been general phase space functions dynamical equations would become intricately coupled and have to be solved numerically as in \cite{bv,dc,js,djs}.

It is straightforward to integrate these Hamilton's equations for $b$, $c$ and $p_c$ variables.  The strategy is to solve the $(c,p_c)$ sector first, then the dynamical equation  for $b$ and finally obtain the solution for $p_b$ using the vanishing of the effective Hamiltonian constraint $H_{\mathrm{eff}} \approx 0$.  The general solution is: 
\ba \label{eq:c}
\tan \Big(\f{\delta_{c}\, c(T)}{2} \Big)&=&  \mp \f{\gamma L_o \dc}{8 m} e^{-2 T},\\
\label{eq:pc} p_c(T) &=& 4 m^2 \Big(e^{2 T} + \f{\gamma^2 L_o^2 \dc^2}{64 m^2} e^{-2 T}\Big) ,
\ea  
\be \label{eq:b}
\cos \big(\delta_{b }\,b(T)\big) = b_o \tanh\left(\f{1}{2}\Big(b_o T + 2 \tanh^{-1}\big(\frac{1}{b_o}\big)\Big)\right),
\ee
where%
\be
b_o = (1 + \gamma^2 \db^2)^{1/2} ,
\ee
and,
\be\label{eq:pb}
p_b(T) = - 2 \f{\sin (\dc\, c(T))}{\dc} \f{\sin (\db\, b(T))}{\db} \f{p_c(T)}{\f{\sin^2(\db\, b(T))}{\db^2} + \gamma^2} .
\ee
 Note that in the classical limit $\db \to 0$ and $\dc \to 0$, these solutions  reduce to (\ref{conf}) and (\ref{momenta}). Next,  (\ref{eq:c}) and (\ref{eq:pc}) immediately imply that\, $p_{c}\sin(\dc c) /(\gamma L_{o}\dc)$  is a Dirac observable which, in the classical limit, has the interpretation of the mass $m$ of the black hole. Since, as we will see,  our effective theory agrees with the classical theory in the low curvature region (e.g. near the black hole horizon for macroscopic black holes) and since  Dirac  observables are constants of motion, we will denote it again by $m$. Thus, in our effective theory:
\be \label{m} m := \Big[ \frac{ \sin\dc c}{\gamma L_{o}\dc}\Big]\, p_{c}, \ee
 which can also be expressed using only $b, p_{b}$ on the constraint surface (see Eq. (\ref{H_eff})):
\be m := - \f{1}{2{\gamma}}\, \Big[\f{\sin \delta_{b} b}{\delta_{b}} + \f{\gamma^{2} \delta_{b}}{\sin\delta_{b}b}\Big]\, \f{p_{b}}{L_{o}} .\ee 
One can pass from the phase space to the space-time description following the same procedure as in the classical theory. Thus, the quantum corrected space-time metric is given by  substituting the lapse $N$ of (\ref{N}) and  solutions (\ref{eq:pc}) and (\ref{eq:pb})  for triads $p_{c}, p_{b}$ in the expression (\ref{metric}).

An important feature of the effective dynamics is that the solutions are non-singular so long as the appropriately chosen quantum parameters $\db$ and $\dc$ are non-zero. In the classical theory, the connection components diverge and the triad components go to zero at the singularity.  This does not occur anywhere in the effective space-time metric.  \emph{In particular, $p_c$ takes a minimum value ${p_c} {\mid}_{\mathrm{min}} = m \gamma L_o \dc$} in every effective space-time. In the phase space picture, the triad  $p_c$ bounces avoiding the central singularity.  This singularity resolution is a direct manifestation of the non-perturbative quantum gravitational effects encoded in the effective Hamiltonian via  quantum parameters $\db$ and $\dc$. 

To summarize, there is a large class of judiciously chosen quantum parameters $\delta_{b}, \delta_{c}$ (discussed in Appendix \ref{a1}) that lead to the quantum corrected, effective space-time geometry given by  Eqs. (\ref{metric}), (\ref{eq:pc}), (\ref{eq:pb}).
 To make a more detailed investigation of properties of the quantum corrected, effective space-time one has to specify $\delta_{b}$ and $\delta_{c}$. We will do this in section \ref{s4}.\\


\section{Causal structure of the effective space-time geometry}
\label{s3}

As we remarked in section \ref{s1}, previous discussions of singularity resolution treated Schwarzschild interior as a cosmological model and focused on issues that are at forefront in anisotropic models, such as bounces of scale factors. Our focus, by contrast, is on black hole aspects. Therefore we will now investigate the consequences of the \emph{phase space} dynamics of section \ref{s2.B} on the \emph{space-time} geometry. Specifically, we will analyze the causal structure in the  extension of Schwarzschild interior provided by the effective metric and show that it is divided by a trapped and anti-trapped regions, separated by a 3-dimensional space-like {\emph{transition surface}} $\T$ that replaces the classical singularity. Results of this section are general in the sense that they are not tied to the specific choice of the quantum parameters introduced in section \ref{s4}; they are consequences of equations of motion (\ref{eq:c}) -- (\ref{eq:pb}) that hold for any $\delta_{b}, \delta_{c}$ in the large family discussed in Appendix \ref{a1}.\medskip

Let us begin by recalling the situation in the \emph{classical theory}. There,  the space-time metric (\ref{metric-kruskal}) corresponding to every dynamical trajectory (with non-zero $m$) admits a black hole (BH) horizon at time $\tau = 2m$, or $T=0$, where the translational Killing vector field $X^{a}$  (with $X^{a}\partial_{a} = \partial/\partial_x$) becomes null and the spatial 3-metric becomes degenerate. In the phase space description, the horizon is characterized by conditions $b=0,\, p_{b} =0$ (see Eqs. (\ref{conf}) and (\ref{momenta})).  At these points the lapse $N_{\rm cl}$ of Eq. (\ref{Ncl})  diverges and so the interpretation in terms of space-time geometry breaks down. Therefore the  horizon represents the past boundary of the interior region. Each dynamical  trajectory also has a future end point at which  $p_{c}$ vanishes ($\tau =0$ or $T =-\infty$)  (see Eq. (\ref{momenta})). In terms of space-time geometry, these points represent the future singularity at which $b, c$ and the Kretschmann scalar diverge.

Let us now examine how this situation changes in the \emph{quantum corrected, effective space-time geometry.}  By construction, the effective metric (\ref{metric}) is again spherically symmetric and has a space-like translational Killing vector field  $X^{a}$. Thus, as in the classical theory, the space-time under consideration is foliated by homogeneous, space-like Cauchy surfaces. The past boundary is again represented by the phase space points $b=0, p_{b} =0$ at which the lapse $N$ of Eq. (\ref{N}) diverges. (Note that, as in the classical theory, along dynamical trajectories vanishing of $b$ implies vanishing of $p_{b}$ and divergence of $N$. See Eqs. (\ref{eq:pb}) and (\ref{N}).) The form (\ref{metric})  of the metric implies that the Killing vector field $X^{a}$ becomes null there.

However, as we already noted in section \ref{s2.B},  Eq. (\ref{eq:pc}) implies that  $p_{c}$ now admits a \emph{non-zero} minimum value, $p_{c}^{\rm min} = \f{1}{2} \gamma\, (L_{o} \delta_{c})\, m$ along every dynamical trajectory. (Recall that $p_{c}$ and $L_{o}\delta_{c}$ are both invariant under the rescalings of the fiducial cell and the fiducial metric.)  As a consequence, none of the curvature scalars diverges: the space-time metric defined by the effective dynamical trajectory is smooth. In the space-time picture, \emph{the 3-surface  $\T$ on which $p_{c}$ achieves its minimum replaces the classical singularity in the quantum corrected geometry}.  To discuss geometrical properties of $\T$,  let us begin by introducing the two future pointing null normals  $\ell^{a}_{\pm}$ to the metric 2-spheres $x= {\rm const}$ and $T= {\rm const}$:
\be
\ell_a^{\pm}=\alpha_\pm\nabla_a T\pm\beta_\pm\nabla_ax.
\ee
The standard normalization conditions 
\be
g^{ab}\ell_a^{\pm}\ell_a^{\pm}=0,\quad g^{ab}\ell_a^{+}\ell_a^{-}=-1, 
\ee
with $\alpha_\pm>0$ and $\beta_\pm>0$, fix three of the parameters $\alpha_\pm$ and $\beta_\pm$ and we will fix the remaining freedom by setting  $\alpha_+=1$. The expansions of these null vectors can be expressed in terms of phase space variables as
\be \label{expn}
\theta_\pm=S^{ab}\nabla_a\ell_b^{\pm}=N^2\, \dot p_c,
\ee  
where $S^{ab}$ is the projection operator on the metric 2-spheres. Since $N$ cannot vanish (see Eq. (\ref{N})), either of the two expansions $\theta_\pm$ vanishes if and only if $\dot p_c=0$, and then they \emph{both} vanish. It follows from (\ref{eq:pc}) that each effective trajectory in the phase space admits one and only one point at which this occurs, and the corresponding time coordinate in the space-time description is given by
\be
T_{\T} = \frac{1}{2} \ln\left(\frac{\gamma L_o \dc}{8 m}\right).
\ee
To the past of the 3-surface $T=T_{\T}$  --\,\,i.e., in the region $ 0 > T >T_{\T}$\,\,-- both expansions $\theta_\pm$ are negative; i.e. the metric 2-spheres are all trapped. To the future of this surface --i.e., for $T_{\T} > T $,  both expansions $\theta_\pm$ are positive; i.e. the metric 2-spheres are all anti-trapped.  (Recall that, by its definition, the coordinate $T$  \emph{decreases} from $T =0$  as we go to the future in the space-time picture and is thus negative in the entire space-time region of interest.) \emph{Therefore ${\T}$ is the transition surface from trapped region to anti-trapped region of the space-time metric (\ref{metric}).}%
\footnote{$\T$ has very interesting geometry. It is a space-like 3-manifold that is foliated by marginally trapped surfaces. However, it is \emph{not} a dynamical horizon because both expansions $\theta_{\pm}$ vanish on $T_{\T}$. Similarly,  although the area of all marginally trapped 2-spheres is the same, $\T$ is \emph{not} a non-expanding  horizon because it is space-like.  These features are quite exceptional: Indeed, we are not aware of any physically interesting space-time in classical general relativity which admits a surface with these interesting properties.}\,
Since $\dot{p}_{c}$ has precisely one zero along each dynamical trajectory, each solution admits one and only one transition surface.   What happens to $\T$ in the classical limit $\db \to 0$ and $\dc \to 0$? In that case $T_{\T} \to -\infty$ which corresponds to the classical black hole singularity. In his precise sense, in the effective description $\T$ replaces the classical singularity.
 
What is the nature of the space-time geometry to the future of the transition surface $\T$? Since both expansions $\theta_{\pm}$ are now positive, the causal structure is completely analogous to the white hole region of  Kruskal space-time. In this sense one can say that $\T$ marks a transition from a black hole interior to the white hole interior. However, we will refrain from using this terminology because to some it suggests that the black hole singularity still persists and the extension corresponds to attaching a white hole geometry to the future of the singularity.  We emphasize that  \emph{the entire geometry is smooth} and $\T$ is invariantly defined as the boundary between a \emph{trapped region} in the past to  an \emph{anti-trapped region} to the future. 

As we saw, the past boundary of the space-time region defined by effective trajectories in our phase space has the interpretation of the black hole horizon since the Killing field $X^{a}$ becomes null there. Since the future of $\T$ represents an anti-trapped region, it is natural to ask if this region also admits a boundary that can be interpreted as the white hole horizon. It follows from the form {of the metric}  (\ref{metric}) that, as in the classical theory, if $p_{b}(T_{0})$ vanishes, then the surface $T=T_{0}$ would represent a Killing horizon.  Eq. (\ref{eq:pb})  implies that this occurs at $T_{0} = - (4/b_{o})\, \tanh^{-1} (1/b_{o})$ because then $\delta_b b(T_{0}) = \pi$ whence $p_{b}=0$. We will see in section \ref{s4} that  for macroscopic black holes this occurs in a low curvature region with our choice of the quantum parameters $\delta_{b},\, \delta_{c}$. Thus, in our effective theory, the `extended Schwarzschild interior' is the smooth space-time region $T_{0}<T<0$ with a black hole type horizon at $T=0$ as its past boundary and a white hole type horizon at $T=T_{0}$ as its future boundary. This portion of the effective space-time is divided into a trapped region to the past of $\T$ and an anti-trapped region to the future of $\T$.\\

\emph{Remarks:} 
\noindent 1. Recall that the transition surface $\T$ in space-time corresponds  to the phase space point at which $p_{c}$ bounces in the corresponding dynamical trajectory. The other phase space momentum variable $p_{b}$ appears only in the expression of  the norm of the translational Killing field $X^{a}$ in space-time. It also undergoes bounces and this generically occurs away from $\T$.  We did not discuss these bounces since these are not significant for   the causal structure of the effective space-time under consideration. 

\noindent 2. The past boundary of the extended Schwarzschild interior is a black hole type (i.e., future) horizon of the classical space-time we started with, while the future boundary is a white hole type (i.e., past) horizon. We will refer to them as `black/white hole type' rather than future/past horizons because in the extended space-time the black hole type future horizon lies to the past of the white hole type past horizon (see Fig. \ref{fig:4}).

\section{Quantum corrected space-time geometry of the Schwarzschild interior}
\label{s4}

This section is organized as follows. In  {section} \ref{s4.A}, we motivate and specify our choice of  quantum parameters $\delta_{b}, \delta_{c}$.  In section \ref{s4.B} we probe the nature of quantum corrections to Einstein's equations that are responsible for the singularity resolution. Even though there is no physical matter anywhere,  it is often convenient to reinterpret non-vanishing of the Einstein tensor in terms of an effective stress energy tensor $\mathfrak{T}_{ab}$ induced by quantum geometry.  We present expressions of  the resulting effective energy density $\rho$ and radial and tangential pressures $\mathfrak{p}_{x}, \, \mathfrak{p}_{\|}$, and show that the strong energy condition is indeed violated by this  $\mathfrak{T}_{ab}$ in a neighborhood of the transition surface $\T$. In this neighborhood, then, there are large departures from classical general relativity. In section \ref{s4.C}  we show that the space-time curvature near $\T$ is of Planck scale. Interestingly, as is common in LQC, each curvature invariant has an \emph{absolute} upper bound, i.e., one that  does not depend on how large the mass is. This is in sharp contrast with the situation in classical general relativity,  where the Kretschmann scalar $K(T) = 48 m^{2}/p_{c}^{3}(T)$ grows with mass at any given $T$, making the `strength' of the central singularity proportional to $m^{2}$. Finally, the effective stress-energy tensor  $\mathfrak{T}_{ab}$ decays away from the transition surface $\T$ and becomes quickly negligible. Thus, for large $m$, Einstein's vacuum equations are satisfied to a high level of accuracy near the black hole and white hole type horizons. The overall situation is similar to that in LQC: quantum geometry corrections are negligible in low curvature regime but grow quickly in the Planck regime, creating an effective repulsive force that resolves the singularity.
Finally, in section \ref{s4.D} we compare and contrast our strategy of fixing $\delta_{b}, \delta_{c}$ and the results that follow  with previous work on the singularity resolution in loop quantization of Kantowski-Sachs model { \cite{ab,lm,bv,dc,dc2,cgp,bkd,cs,js,djs,oss,cctr}.}



\subsection{Transition surface, area gap and $\delta_{b},\delta_{c}$}
\label{s4.A}

As indicated in section \ref{s1},  several different choices of quantum parameters $\delta_{b}, \delta_{c}$ have been made in the literature  \cite{ab,lm,bv,dc,cgp,bkd,cs,djs,oss,cctr}, including those where $\delta_b$ and $\delta_c$ are not constants \cite{bv,dc,bkd,djs,cctr}, leading to quite different effective descriptions of Schwarzschild interior. In this sub-section, we will first motivate and then specify our choice. In section \ref{s4.D} we will compare and contrast the physical predictions that result from different choices.

Recall  from section \ref{s2} that the gravitational connection $A_{a}^{i}\tau_{i}$  enters in the Hamiltonian constraint via its curvature $F_{ab}^{i}\tau_{i}$.  In the passage to quantum theory,  there is a surprising result: the requirement of background independence selects a unique representation of the canonical commutation relations  (in full LQG \cite{lost,cf}, as well as in LQC \cite{aamc,eht}).  In this representation, there is no operator corresponding to the connection $A_{a}^{i}\tau_{i}$ itself; only the operators corresponding to the holonomies $h_{\ell}$ defined by $A_{a}^{i}\tau_{i}$ along links $\ell$ are well-defined. Therefore, in the quantum theory components of the curvature $F_{ab}^{i}\tau_{i} $ have to be expressed using holonomies (see, e.g., \cite{alrev,apbook}). In the classical theory, one can calculate, say, $F_{\theta,\phi}^{i}\tau_{i}$  as follows: first evaluate the ratio \,\,$\big( h_{\Box (\theta,\phi)} -1\big) /\big( {\rm Ar}({\Box (\theta,\phi)})\big)$\, --where $h_{\Box(\theta,\phi)}$ is the holonomy around a closed rectangular  plaquette $\Box(\theta,\phi)$ within the $\theta$-$\phi$ 2-sphere enclosing an area ${\rm Ar}({\Box (\theta,\phi)})$--\, and then take the limit as the plaquette  $\Box(\theta,\phi)$ shrinks to a point. The idea is to use this procedure also in the quantum theory. However,   the area operator has a discrete spectrum in LQG and there is a minimum non-zero area eigenvalue --the area gap $\Delta$. Therefore, the strategy is to obtain the quantum operator corresponding to the classical curvature component $F_{\theta,\phi}^{i}\tau_{i}$ by shrinking the plaquette $\Box(\theta,\phi)$  only till its area ${\rm Ar}({\Box (\theta,\phi)})$ equals $\Delta$. In the same manner, operators corresponding to the other two curvature components $F_{\phi, x}^{i}\tau_{i}$ and $F_{\theta, x}^{i}\tau_{i}$ are defined as holonomies along plaquettes $\Box(\phi, x)$ and $\Box(\theta, x)$ in the $\phi$-$x$ and $\theta$-$x$ planes enclosing area $\Delta$.   Therefore, we have:
\be \hat{F}_{ab}^{i}\, \tau_{i} = \f{1}{\Delta}\,\,  \big(h_{\Box_{ab}}  - 1\big), \ee
where the appropriately chosen plaquette $\Box_{ab}$ lies in the $a$-$b$ plane, enclosing area $\Delta$. Consequently  the operator corresponding to curvature now acquires a Planck scale non-locality which lies at the heart of quantum corrections to dynamics that naturally resolve singularities.

The quantum parameter $\delta_{b}$ has the interpretation of the length of  each link constituting the plaquette within the $\theta$-$\phi$ 2-spheres, and $\delta_{c}$, of the length of the links in the $x$-direction within the plaquettes in the $\theta$-$x$ and $\phi$-$x$ planes in the fiducial cell $\mathcal{C}$.  First investigations \cite{ab,lm} of the Schwarzschild interior followed the procedure  initially used in FLRW models \cite{abl}   --known as the $\mu_{o}$-scheme-- and set  $\delta_{b}, \delta_{c}$ to a constant, $\delta$ (see footnote \ref{fn1})).  Later investigations revealed that the resulting quantum dynamics has several limitations \cite{js,cs}. (For example, its physical predictions depend on the choice of fiducial structures.)  These were overcome in  an `improved dynamics' scheme in {Ref. \cite{bv} (and a variant in Ref. \cite{dc})} by mimicking the successful `$\bar\mu$-scheme' introduced for the  FLRW models in Ref. \cite{aps3}. Then, $\delta_{b}, \delta_{c}$ turn out to be specific functions on the phase space \emph{whose values evolve along the effective dynamical trajectories}. However, as we discuss in section \ref{s4.D}, effective theories based on {all of} these choices of $\delta_{b}, \delta_{c}$ have physically unacceptable features.  We therefore follow a procedure that straddles between the $\mu_o$ and $\bar \mu$ schemes:  As mentioned before, our $\delta_{b}, \delta_{c}$ will not be constants all over the phase space, but  they will be constants along dynamical trajectories (as in \cite{cs}). That is, they will Poisson-commute with the effective Hamiltonian constraint.
 
Our strategy is to  fix the Dirac observables $\delta_{b}, \delta_{c}$  by demanding  that the plaquette $\Box(\theta,\phi)$ and $\Box(\phi, x)$ should enclose minimum area when evaluated on $\T$. (By spherical symmetry, the condition is then satisfied also for the plaquette $\Box(\theta,x)$.)    Our $\delta_{b}, \delta_{c}$ will then be well-defined Dirac observables because each effective trajectory admits \emph{one and only one} point at which $\dot{p}_{c} =0$ (which, in the effective space-time geometry defines the transition surface $\T$; see Eq. (\ref{expn})).  Now, since the parameters $\delta_{b}, \delta_{c}$ used in the $\mu_{o}$-type scheme \cite{ab,lm} are constants on the entire phase space, they are also  (trivially) Dirac observables. Our procedures differs from the $\mu_{o}$ scheme because we evaluate the area using the \emph{physical} effective metric --rather than the fixed fiducial metric used in \cite{ab,lm}-- making a crucial use of the transition surface $\T$. Therefore our $\delta_{b}, \delta_{c}$ are not constants on the phase space but \emph{vary from one effective dynamical trajectory to another.} 

Let us begin with an infinitesimal rectangular plaquette $\Box(\phi, x)$ in the $\theta=\pi/2$ plane of our fiducial cell. The plaquette has two parallel links along the $x$-axis and two parallel links along $\theta=\pi/2$. Let $\delta_{c}$ denote the \emph{fractional} length of the link along $x$ axis. Note that fractional lengths are metric independent. For example, with respect to the fiducial metric ${\fq}$ of  Eq. (\ref{fiducial}), the total length of the fiducial cell $\mathcal{C}$ along the $x$-direction is $L_{o}$, and the length of our link will be $\delta_{c}L_{o}$. Similarly, with respect to the physical metric, the total length of the fiducial cell along $x$-direction is $(|p_{b}|/\sqrt{p_{c}})$ and the length of our link will be $(|p_{b}|/\sqrt{p_{c}})\, \delta_{c}$. Likewise, let the \emph{fractional} length of the link along the equator be $\delta_{b}$. Then,  from the form  (\ref{metric}) of the physical metric, we conclude that  the \emph{physical} area enclosed by the plaquette $\Box(\phi, x)$ at the transition surface $\T$ is:
\be \label{area1}  {\rm Ar} (\Box(\phi, x))  = \delta_{b}\,\delta_{c}\, (2\pi |p_{b}|\big{|}_{\T}).  \ee
Since the total area $A_{\phi,x}$ of the $\phi$-$x$ plane in the fiducial cell is $2\pi |p_{b}|_{\T}$, as expected  $\delta_{b}\delta_{c}$ has the invariant  interpretation as the \emph{ratio} of the area ${\rm Ar} (\Box(\phi, x))$ enclosed by the plaquette $\Box(\phi, x)$ and the total area of the $\phi,x$-plane within the fiducial cell. We discussed these geometric properties in some detail to distinguish the present scheme from others in the literature. There, $\delta_{b}, \delta_{c}$ are generally taken as coordinate lengths using $\theta, \phi, x$ and so their invariant geometrical meaning remains unclear.%
\footnote{Also, this careful analysis is essential to get the correct  $2\pi$ and $4\pi$ type numerical factors in the expressions of $\delta_{b},\delta_{c}$ in Eq. (\ref{db-dc}). Some of the physical properties depend on these factors. For example, in a less careful treatment that ignores these factors,  the mass $m$ changes as one moves from one asymptotic region to another one to its future (e.g., from region I to III in Fig. \ref{fig:4}), even in the large mass limit.}

Next, let us consider the plaquette $\Box(\theta,\phi)$ in any $x={\rm const}$ 2-sphere on the transition surface $\T$. Because the 2-spheres are round, we are led to use the same fractional length $\delta_{b}$ along the two orthogonal directions of the plaquette. Then it follows from the form (\ref{metric}) of the metric that the \emph{physical} area enclosed by this plaquette on the transition surface $\T$ is
\be \label{area2}  {\rm Ar} (\Box(\phi, x))  = (\delta_{b})^{2}\, (4\pi p_{c} \big{|}_{\T}),  \ee
so that now $(\delta_{b})^{2}$ has  the interpretation of the ratio of the area enclosed by the plaquette to the total area of the 2-sphere.

We can now implement the main strategy: We will constrain $\delta_{b}, \delta_{c}$ by  requiring that the areas enclosed by the two plaquettes on the transition surface $\T$ be equal to the area gap:
\be\label{area3}
2\pi\,\dc\db\, |p_b| \big{|}_{\T} =\Delta, 
\ee
and
\be\label{area4}
4\pi\, \db^2\, p_c\big{|}_{\T}=\Delta. 
\ee
Since on each dynamical trajectory $p_{b}$ and $p_{c}$ have fixed values on the transition surface, it is intuitively clear that the two equations  would determine the values of $\delta_{b}, \delta_{c}$. This is indeed the case under the well-motivated assumptions $\delta_{b} \ll 1,\, \delta_{c} \ll 1$ and $m \gg \lp$  (where $m$ is the phase space function defined in Eq. (\ref{m})). Since the proof of this result  is rather technical and requires a significant detour, to  maintain the flow of the argument we present it in Appendix \ref{a2}. The final result is that in the large $m$ limit, we have: 
\be\label{db-dc}
\db=\Big(\frac{\sqrt{\Delta}}{\sqrt{2\pi}\gamma^2m}\Big)^{1/3}, \qquad 
L_{o}\dc=\frac{1}{2} \Big(\frac{\gamma\Delta^2}{4\pi^2 m}\Big)^{1/3}.
\ee
(Recall that it is only $L_{o} \delta_{c}$ that has invariant meaning in the sense of being independent of the choice of the fiducial metric and cell).  Note that both parameters depend on mass and go as $m^{-\f{1}{3}}$. This property is important for physical properties of the resulting effective metric.

\subsection{Quantum corrections to Einstein's equations}
\label{s4.B}

From the perspective of classical general relativity it is natural to investigate how the effective theory manages to resolve the Schwarzschild singularity.  The effective stress energy tensor induced by quantum corrections,
\be
\mathfrak{T}_{ab} :=\frac{1}{8\pi G}\,\,G_{ab},
\ee
must violate standard energy conditions. It is natural to ask: How large are the violations? and, Where do they occur? We will now discuss these issues.  

Let us begin by noting that  $\mathfrak{T}_{ab}$ can be interpreted as the stress energy tensor of an anisotropic perfect fluid with effective energy density 
\be \label{rho}
\rho=-\mathfrak{T}^0_0=\frac{1}{8\pi G}\left(\frac{1}{p_c}+\frac{1}{N^2}\frac{\dot{p}_b\dot{p}_c}{p_b p_c}-\frac{1}{N^2}\frac{\dot{p}_c^2}{4 p_c^2}\right),
\ee
and radial and tangential pressures 
\ba 
\mathfrak{p}_x&=&\mathfrak{T}^1_1 = \frac{1}{8\pi G}\left(-\frac{1}{p_c}+\frac{1}{N^2}\frac{\dot{p}_c^2}{4 p_c^2}-\frac{1}{N^2}\frac{\ddot{p}_c}{p_c} +\frac{\dot{N}}{N^3}\frac{\dot{p}_c}{p_c}\right),\\
\mathfrak{p}_{\|} &=&\mathfrak{T}^2_2 = \frac{1}{8\pi G}\left(\frac{1}{N^2}\frac{\dot{p}_b\dot{p}_c}{2 p_b p_c}-\frac{1}{N^2}\frac{\dot{p}_c^2}{4p_c^2}-\frac{1}{N^2}\frac{\ddot{p}_b}{p_b}+\frac{\dot{N}}{N^3}\frac{\dot{p}_b}{p_b}\right). \label{p}
\ea
Since $\dot{p}_{c} =0$ at the transition surface $\T$, we have
\be \label{violation}
\left.(\mathfrak{T}_{ab}-\frac{1}{2}g_{ab}\mathfrak{T})T^aT^b\right|_{\T} =  \left.\left(\rho+\mathfrak{p}_x+2\mathfrak{p}_{\|}\right)\!\right|_{\T}=  \f{1}{8\pi G}\,\left.\left(\frac{\dot N}{N^3}\frac{2\dot{p}_b}{p_b}-\frac{1}{N^2}\frac{2\ddot{p}_b}{p_b}-\frac{1}{N^2}\frac{\ddot{p}_c}{p_c}\right)\right|_{\T},
\ee
Note that the right hand sides of (\ref{rho}) - (\ref{p}) hold for any choice of lapse. Therefore one can evaluate them using our choice (\ref{N})   (using $\delta_b, \delta_c$ as in 
(\ref{db-dc})). We find $\ddot{p}_c >0, (\ddot{p}_b/p_{b}) >0$, and $\dot{p}_b$ is much smaller than other terms in the expression, making the right side of (\ref{violation}) negative. (Indeed, as the plots in Figs. \ref{fig:2} show,  $\rho, \mathfrak{p}_{x}$ and $\mathfrak{p}_{\|}$ are all negative at the transition surface for our choice of $\delta_{b}, \delta_{c}$, whence the middle term is manifestly negative.) Therefore, we conclude that
\be
\left.(\mathfrak{T}_{ab}-\frac{1}{2}g_{ab}\mathfrak{T})T^aT^b\,\right|_{\T}<0.
\ee
Thus, for macroscopic black holes considered in this paper, the strong energy condition is violated at (and therefore in a neighborhood of)  $\T$, just as one would expect.

\subsection{Universal upper bounds on curvature invariants}
\label{s4.C}

The explicit solutions to Hamilton's equations given in section \ref{s2.B} show that the phase space variables are manifestly finite along effective dynamical trajectories. Therefore it is clear that the space-time metric (\ref{metric}) is smooth throughout the open interior region bounded by the two horizons.  Therefore, in any one effective solution, curvature scalars are all finite and therefore bounded above.  However,  these upper bounds could well diverge in the limit $m\to \infty$. \emph{Interestingly, this does not happen: each curvature invariant has an absolute, finite upper bound in this limit.}  Existence of such \emph{uniform} upper bounds appears to be a general occurrence  in LQG.  It could be a reflection of a deeper property of quantum geometry effects at the heart of the mechanism that leads to the resolution of strong, space-like singularities in LQG {\cite{ps,ss}.}

\nopagebreak[3]\begin{figure} [h] 
\includegraphics[width=0.49\textwidth]{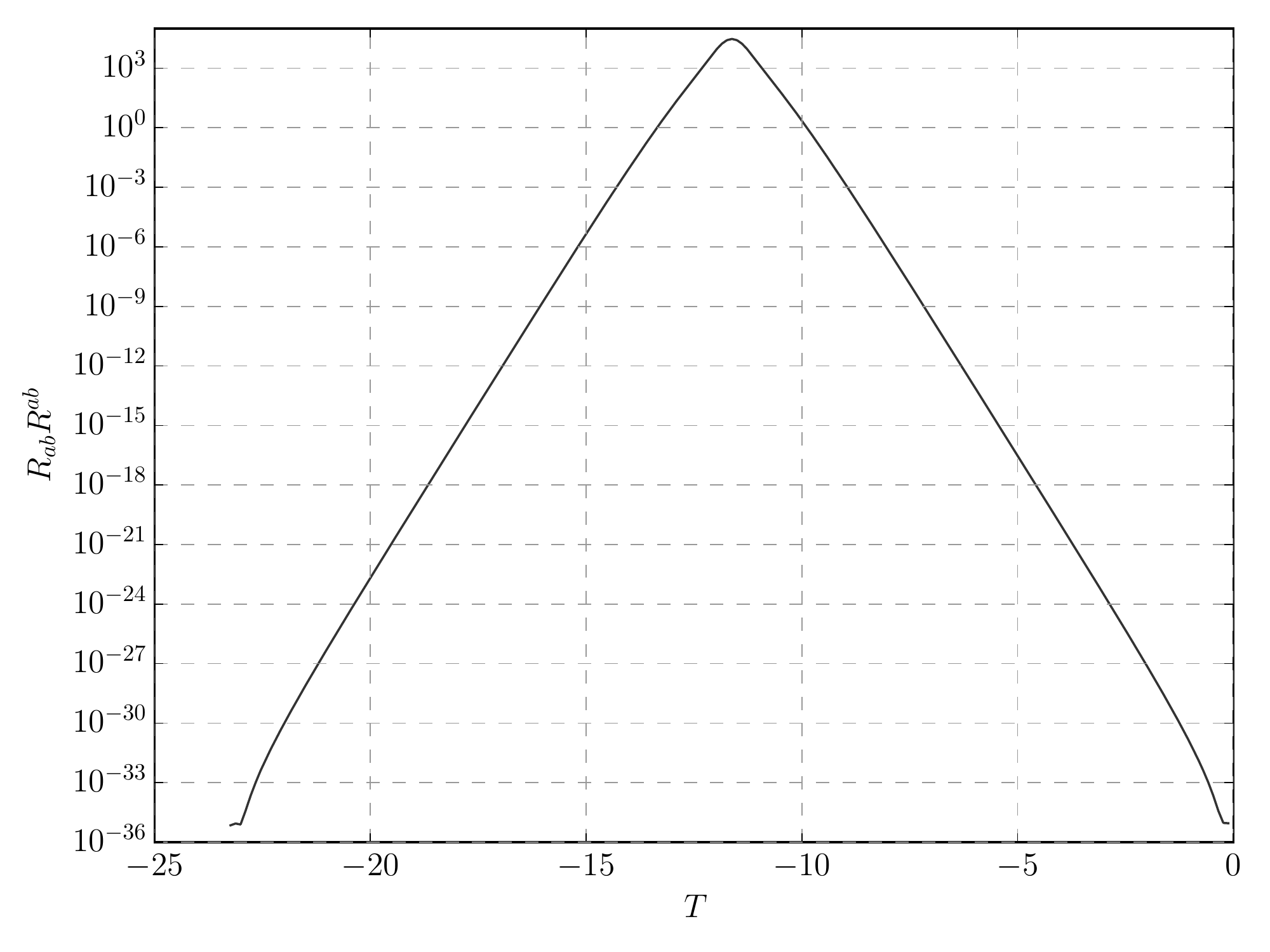}  \hskip0.2cm
\includegraphics[width=0.49\textwidth]{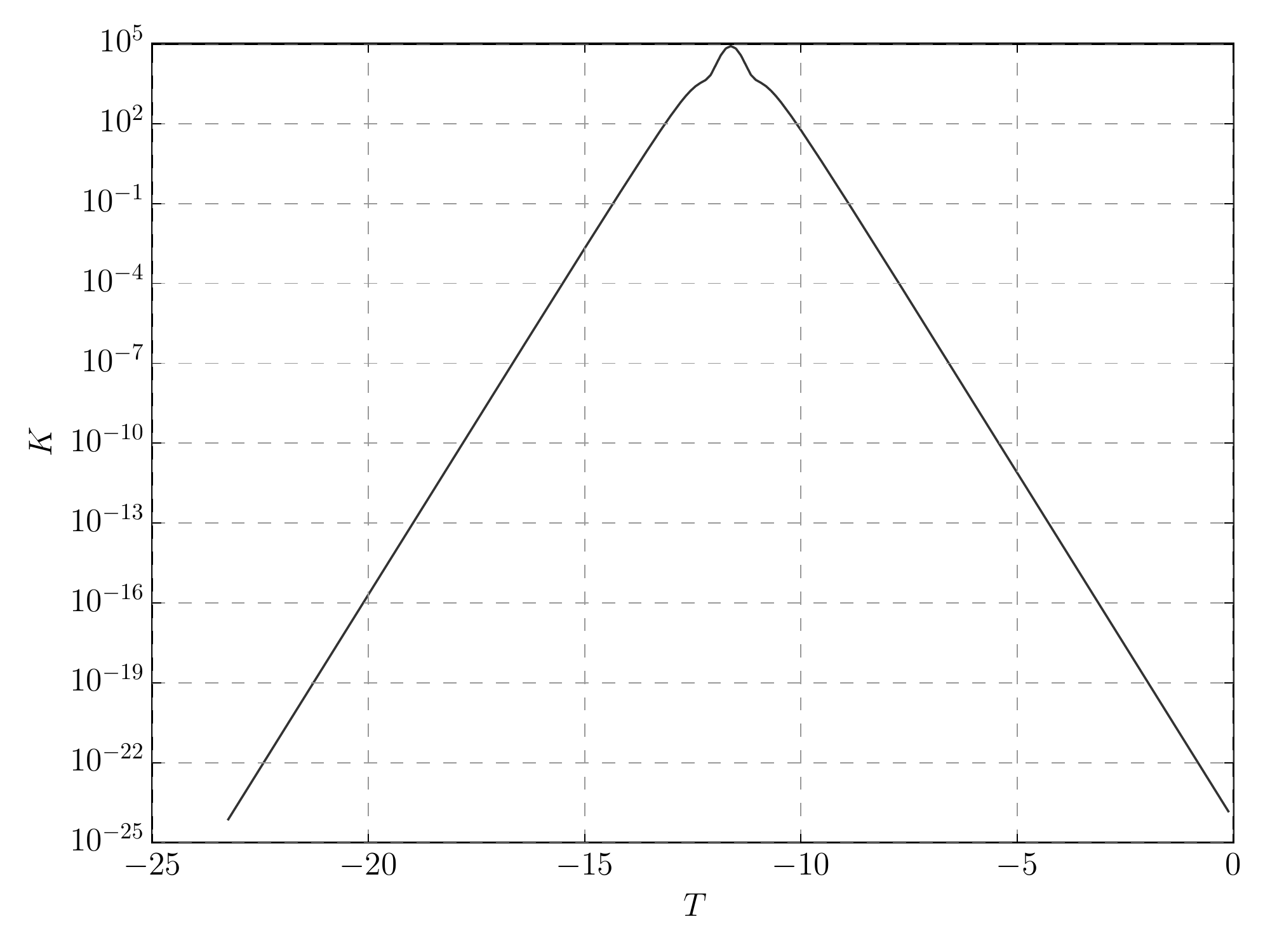}
\caption{\footnotesize{
Time evolution of curvature scalars in the quantum corrected Schwarzschild interior for $m = 10^{6}$ { (in Planck units)}.  (The time parameter $T$ is negative in the space-time region under consideration and decreases as we move to the future.) The black hole type horizon occurs at $T=0$, the transition surface $\T$ lies at $T= -11.62$ and the white hole type at $T= -23.24$. Space-time region to the past of $\T$ is trapped and to the future of $\T$ is anti-trapped.  Curvature scalars are bounded throughout  this evolution and attain their only maximum on $\T$ that replaces the classical singularity. \emph{Left Panel:} The invariant $R_{ab}R^{ab}$. The Ricci curvature is induced by quantum corrections and responsible for the singularity resolution. Although it is of Planck scale near $\T$, it decays rapidly away from $\T$ and is of the order of $10^{-35}$ in Planck units near the two horizons. \emph{Right Panel:} The Kretschmann scalar $R_{abcd} R^{abcd}$. It also has a single maximum at $\T$, decreases as we move away from $\T$  and is extremely close to the classical value $3/(4m^{4}) \approx 10^{-24}$ in Planck units near the two horizons. Thus, the ratio $R_{ab}R^{ab}/K_{\rm class}$ is very small,\, $\sim  10^{-13}$\, near the horizon even when the black hole has as small a mass as $10^{6} M_{\rm Pl}$  and it becomes much smaller for truly macroscopic black holes. }}
\label{fig:1}
\end{figure}

Since we know the explicit time dependence  of the phase space variables from Eqs. (\ref{eq:c}) -- (\ref{eq:pb}),   using the form (\ref{metric}) of the space-time metric we can calculate various curvature scalars at the transition surface. We used \texttt{MATHEMATICA} to simplify these expressions in the large $m$ limit. The results can be summarized as follows: At the transition surface $\T$,
\begin{itemize}
\item the (square of the) Ricci scalar has the asymptotic form:
\be
R^{2}\mid_{\T}\,\,=\,\,\frac{256\pi^{2}}{\gamma^4\Delta^{2}}+{\cal O}\Big(\big(\f{\Delta}{m^2}\big)^{\f{1}{3}}\,\ln \frac{m^2}{\Delta}\Big); 
\ee
\item the square of the Ricci tensor has the asymptotic form
\be
R_{ab}R^{ab}\mid_{\T}\,\,=\,\,\frac{256\pi^2}{\gamma^4\Delta^2}+{\cal O}\Big(\big(\f{\Delta}{m^2}\big)^{\f{1}{3}}\,\ln \f{m^2}{\Delta} \Big);
\ee 
\item the square of the Weyl tensor has the asymptotic form
\be
C_{abcd}C^{abcd}\mid_{\T}\,\,=\,\, \frac{1024\pi^2}{3\gamma^4\Delta^2}+{\cal O}\Big(\big(\f{\Delta}{m^2}\big)^{\f{1}{3}}\, \ln \f{m^2}{\Delta}\Big);
\ee 
\item and, consequently, the Kretschmann scalar $K = R_{abcd}R^{abcd}$ has the asymptotic form 
\be
K\mid_{\T}\,\, =\,\, \frac{768\pi^2}{\gamma^4\Delta^2}+{\cal O}\Big(\big( \f{\Delta}{m^2}\big)^\f{1}{3} \,\ln\f{m^2}{\Delta}\Big).
\ee 
\end{itemize}

These expressions have two notable features. First, the area gap $\Delta$ appears in the denominator, bringing out the fact the finiteness of all upper bounds can be directly traced back to quantum geometry.  Second, the leading terms are \emph{mass independent} and their denominator is quadratic in $\gamma^{2}\Delta$   (which, by footnote \ref{fn4} equals $\Delta^{3}/(48\pi^{2}\lp^{4})$). However, the numerical coefficients vary. (The same pattern is encountered in LQC of the FLRW models.)

\begin{figure} [h]
\begin{center}
\includegraphics[width = 0.49\textwidth]{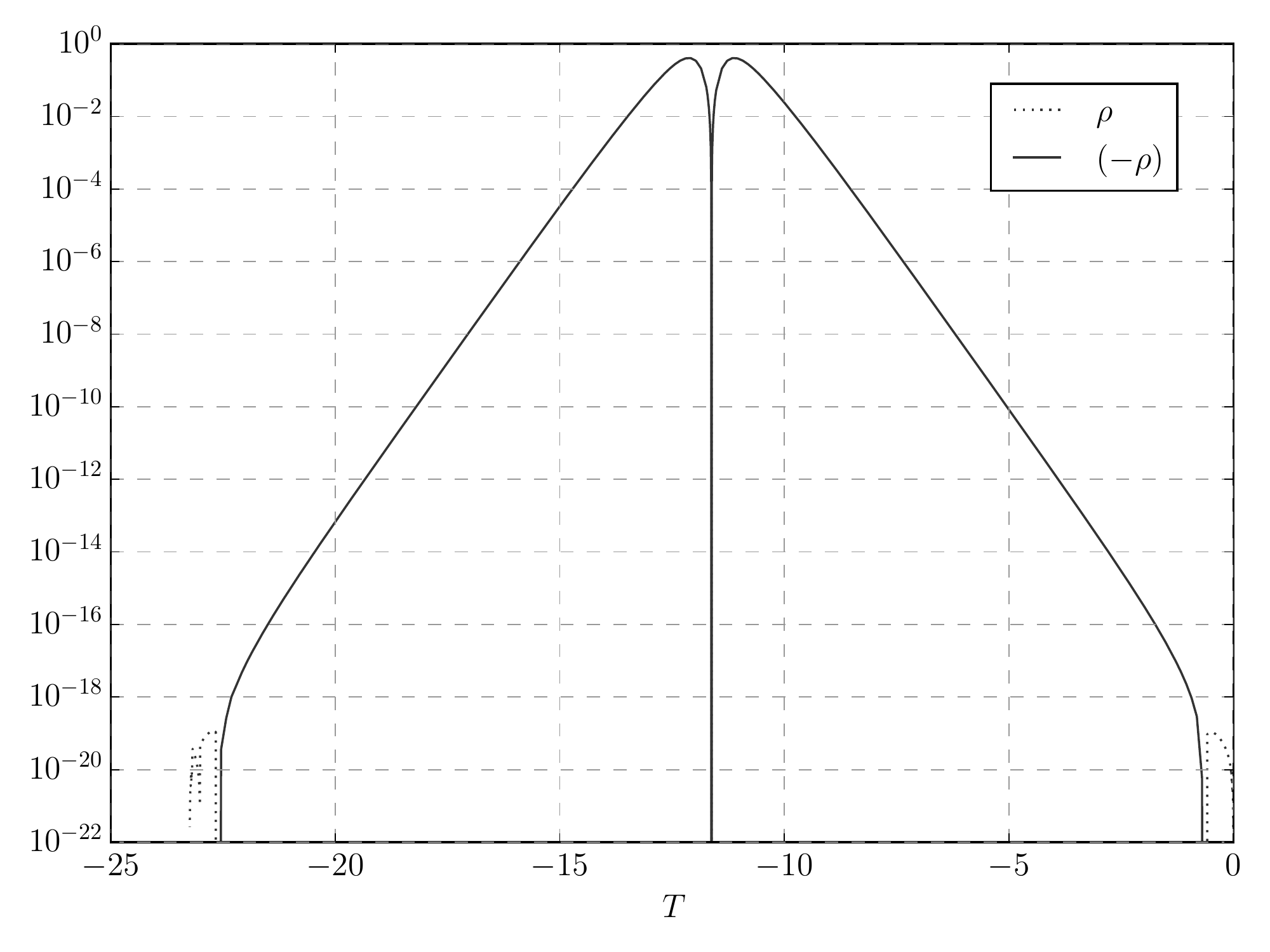}  
\includegraphics[width = 0.49\textwidth]{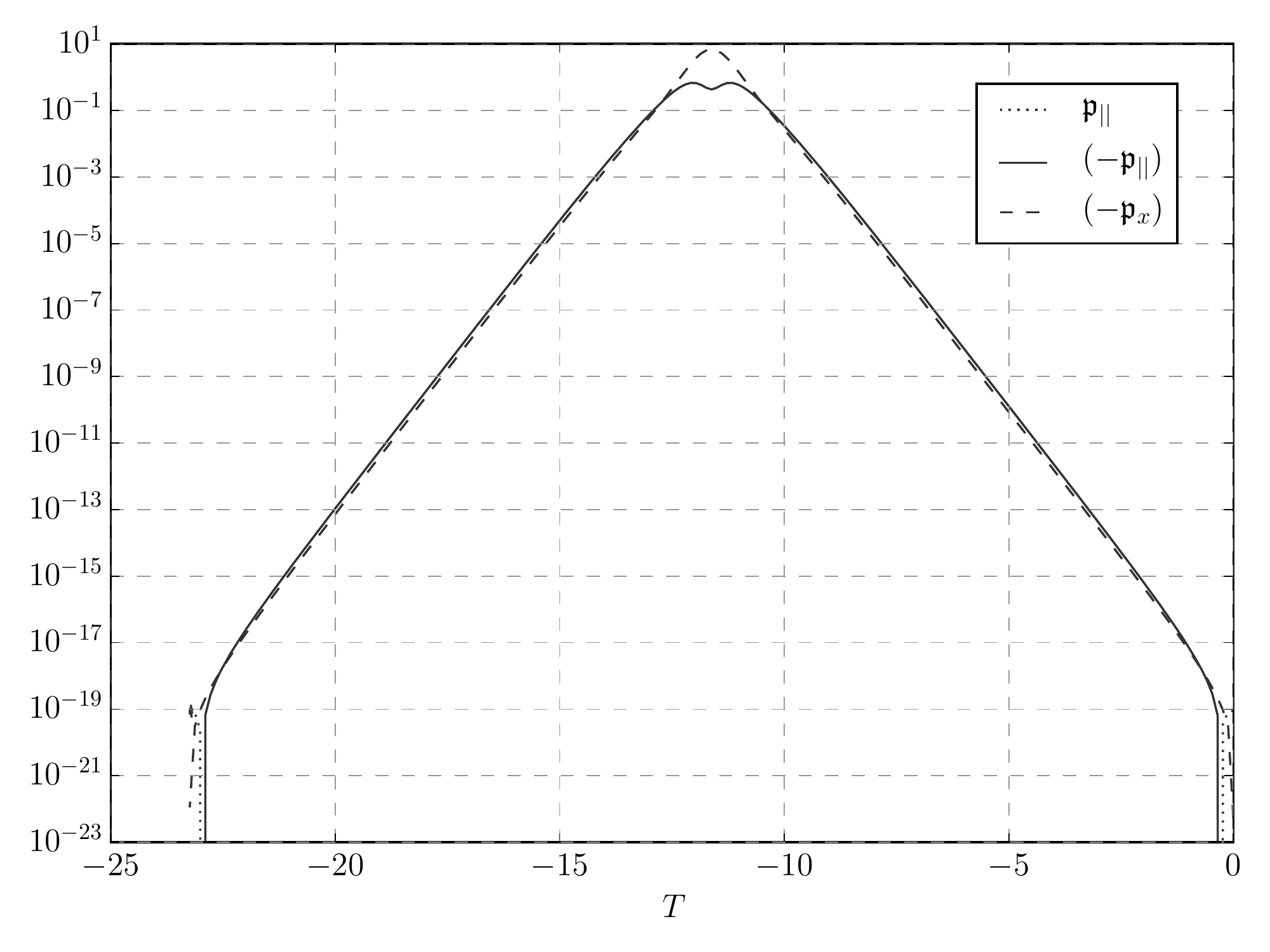} 
\caption{\footnotesize{Time evolution in the quantum corrected Schwarzschild interior for the same mass as in Fig.\ref{fig:1} ($m = 10^{6}$). All quantities plotted are identically zero in classical general relativity and have their origin in quantum geometry. They attain their only maximum at the transition surface and decay rapidly away from $\T$.  \emph{Left Panel:} energy density $\rho$ is negative almost everywhere (solid line) in the `interior region' except in small neighborhoods of the two horizons (dotted lines). \emph{Right Panel:} radial pressure $\mathfrak{p}_{x}$ (dashed line)  and tangential pressure $\mathfrak{p}_{\|}$ (solid line) are both negative almost everywhere in the interior region, but the tangential pressure $\mathfrak{p}_{\|}$ becomes positive in small neighborhoods of the two horizons (dotted lines).}}
\label{fig:2}
\end{center}
\end{figure}

Note that these asymptotic forms refer to the transition surface $\T$. Since there is a precise sense in which the classical singularity is replaced by $\T$ in the effective theory, intuitively it is clear that these values would also be the upper bounds on curvature scalars of the effective metric throughout the space-time region under consideration. However, since the expressions of these scalars at a general time are much more intricate, it is difficult to verify the validity of this expectation analytically. (For example, while the expression of the Kretschmann $K$ scalar  is simply $48 m^{2}/ p_{c}^{3}$ in classical general relativity, it has  more than twenty complicated terms in the effective theory.)   Therefore we carried out numerical evaluations for several values of the mass parameter $M = m/G$. Figs \ref{fig:1} and \ref{fig:2}  illustrate the situation for $m=10^{6} \lp$.  The Ricci tensor $R_{ab}$, the energy density $\rho$ and the effective pressures $\mathfrak{p}_{x}, \mathfrak{p}_{\|}$ are all zero in classical general relativity.  But they acquire large Planck scale values near the transition surface $\T$ which, however,  decay very rapidly as we move away from $\T$.  Near the two horizons, their values are  $\sim 10^{-20}$  or less,  while $K^{1/2}$ --the square-root  of the  Kretschmann scalar which has the same dimensions-- is of the order $10^{-12}$ there. Thus, the contribution of the Ricci tensor to the total curvature is completely negligible near the horizon already for black holes whose Schwarzschild radius is as small  as $10^{6}\lp$ and they become even more negligible for truly macroscopic black holes. Thus, just as in LQC,  although the quantum geometry corrections are sufficiently large in the Planck regime to resolve the singularity, they decay rapidly as space-time curvature becomes a few orders of magnitude smaller. In this precise sense,  quantum gravity corrections play no role near horizons of large black holes in our model, contrary to what is sometimes suggested in other programs (see, e.g., \cite{sg}). Finally,  Fig. \ref{fig:2}  shows that $\rho, \mathfrak{p}_{x}, \mathfrak{p}_{\|}$ are all negative in a large neighborhood of $\T$, whence the strong energy condition is violated there, just as one would expect from the singularity resolution. 

\subsection{Comparison with prior LQG investigations}
\label{s4.D}

\begin{figure} [h] 
\includegraphics[width=0.49\textwidth]{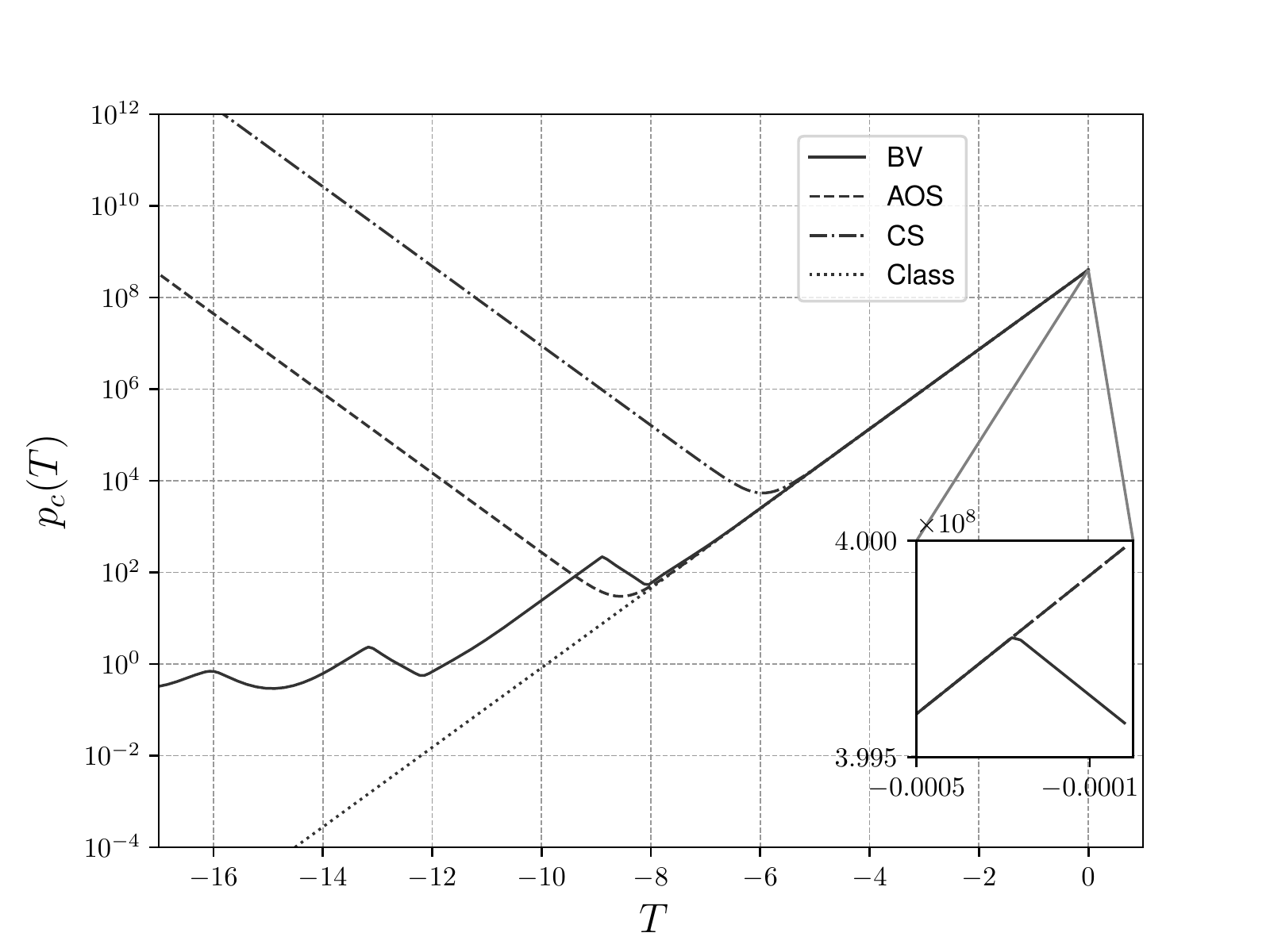}  
\hskip0.2cm
\includegraphics[width=0.49\textwidth]{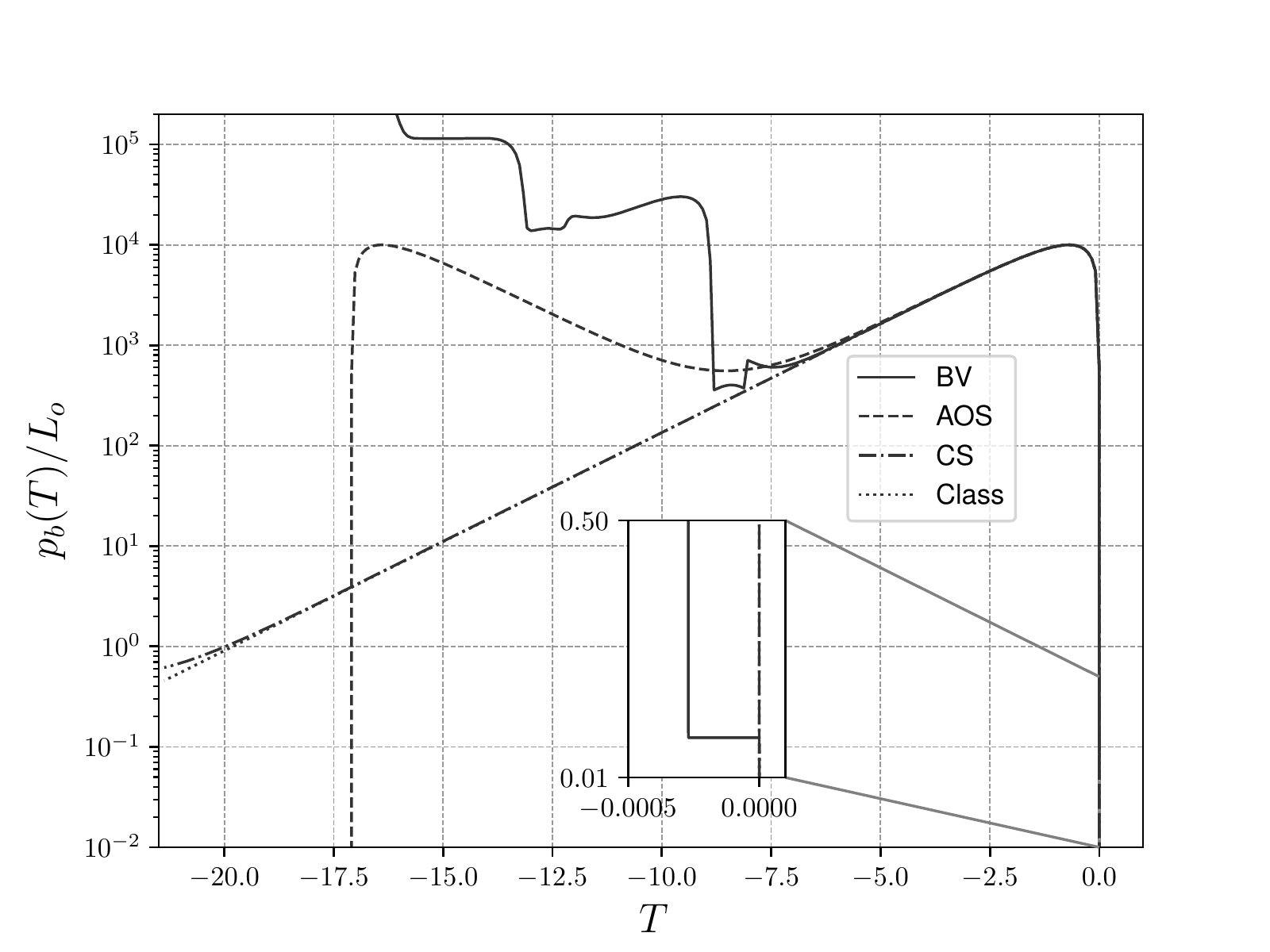}
\caption{\footnotesize{Comparison between the dynamical behavior of the triad components $p_{c}$ and $p_{b}$ in various LQG approaches for $m=10^{4}$. In the effective space-time geometry, transition surfaces $\T$ occur each time $p_{c}$ undergoes a bounce. They separate trapped and anti-trapped regions.  (As in previous figures, the black hole type horizon lies at $T=0$ and $T$ becomes more and more negative as time evolves.)  The label `Class' refers to classical dynamics in which there is no bounce; the label `CS' refers to the `generalized' $\mu_{o}$ scheme \cite{cs} discussed in section \ref{s4.D.3}; `AOS' refers to the dynamics in our approach discussed in sections \ref{s3} and \ref{s4.A} - \ref{s4.C}; and `BV' refers to dynamics in the $\bar\mu$-type scheme \cite{bv} discussed in section \ref{s4.D.2}.
\emph{Left Panel: Evolution of $p_{c}$.} In the classical theory $p_{c}$ decreases steadily corresponding to the monotonic decrease in the radius of the round 2-spheres. In `CS' and `AOS', it undergoes precisely one bounce, with trapped region to the past of the bounce and anti-trapped to the future. In BV it undergoes several bounces. The anti-trapped region after the first bounce is very short lived. After the second bounce this $\bar\mu$ scheme cannot be trusted because its underlying assumptions are violated. \emph{Right Panel: Evolution of $p_{b}$.} This triad component does not play a direct role in determining the  trapped and anti-trapped regions. But it enters in the expression of the norm of the translational Killing field $X^{a}$ and its vanishing signals the emergence of a black hole or white hole type horizon. The white hole type horizon  emerges much later in the CS approach  \cite{cs} than in AOS  reflecting the fact that there is a  large mass amplification in the CS approach while there is no amplification in the AOS approach (in the large $m$ limit). The BV approach   \cite{bv}  becomes unreliable after $T \sim -12$. The zooms shows another limitation of the BV approach: very near the black hole type horizon, the BV dynamics deviates from the classical theory even though space-time curvature is still small. The AOS dynamics is indistinguishable from classical dynamics near this horizon.
}}
\label{fig:3}
\end{figure}

As mentioned  in section \ref{s1}, previous LQG investigations of the Schwarzschild interior using Kantowski-Sachs cosmology can be naturally divided into three classes.  We will compare and contrast our strategy and results with those used in these three types of schemes. Some key differences predicted by various approaches are shown in the dynamics of $p_{c}$ and $p_{b}$ in Fig. \ref{fig:3}.

\nopagebreak[3]

\subsubsection{$\mu_{o}$-type approaches}
\label{s4.D.1}
Ref. \cite{abl} used quantum kinematics that descends from full LQG and showed that the big-bang singularity in FLRW models is naturally resolved, thanks to the area gap in LQG. This strategy has since come to be known as the $\mu_{o}$ scheme. The underlying ideas were  carried over to the analysis of the Schwarzschild singularity in Refs. \cite{ab,lm}.  The kinematical framework introduced in \cite{abl,ab}  and the idea of incorporating quantum gravity corrections to dynamics by representing curvature in terms of holonomies around `elementary plaquettes' continue to be widely used in the analysis of cosmological and black hole singularities. However, subsequent investigations of detailed predictions brought out the fact that the specific implementation of this strategy in \cite{ab,lm}  has several important drawbacks (see, e.g., \cite{aps2,aps3,js,cs}).  In this sense,  while investigations like those in Refs. \cite{abl,ab}  served to open a fruitful avenue, they have to be suitably modified for physical viability.

In the Schwarzschild case, the situation can be summarized as follows.  In Ref \cite{ab,lm,cgp}, the new quantum parameters $\db,\dc$ are assumed to be constants: the `area-gap argument' was used to set  their values to
\be \label{munot}
\db = \dc  = 2\sqrt{3} =:\delta. 
\ee 
(More precisely, the fractional length of the link in the $x$ direction was taken to be $\delta$ and the coordinate lengths in the $\theta$ and $\phi$ direction were taken to be $\delta$.)
Constancy of these parameters simplifies the analysis considerably and it is possible to obtain the explicit action of the Hamiltonian constraint operator on the kinematical Hilbert space.  In the classical theory, $p_{c}$ can be interpreted as an internal time. This choice is viable since (up to a factor of $4\pi$) it determines the area of 2-spheres, which in the space-time language equals $\tau^{2}$.  The form of the constraint operator is such that the quantum constraint  equation can be thought of as providing an evolution along the `internal time variable' provided by eigenvalues of $\hat{p}_{c}$. One can then verify that the singularity is absent in the quantum evolution.

To understand this prediction in detail,  dynamics of this model was analyzed in detail in \cite{cs} using effective field equations.%
\footnote{Although the effective equations are yet to be systematically derived from the quantum evolution in this model, experience with anisotropic cosmological models \cite{numlsu-4}  suggests that for macroscopic black holes they will approximate the exact evolution quite accurately if the quantum state is chosen to be sharply peaked along the classical dynamical trajectory initially, i.e. in the weak curvature region.}
The key result on singularity resolution holds also in the effective theory and, furthermore,  one now has a detailed understanding of the quantum corrections to Einstein's equations that are responsible for this resolution. In particular,  the area of the round 2-spheres --encoded in $p_{c}$-- decreases to a minimum \emph{non-zero} value and then increases again till one arrives at a white-hole type horizon. However, the analysis also revealed a key  limitation of the way in which the main ideas are implemented in \cite{ab,lm}.  It turns out that physical quantities such as values of expansion and shear  (of the normal to the homogeneous slices), as well as the minimum value of $p_{c}$ \emph {depend on the value of $L_{o}$ used to construct the fiducial cell $\mathcal{C}$}. Another key observable is the radius of the white hole type horizon that determines as the ADM mass in the corresponding asymptotic region. This too depends on $L_{o}$. Thus, while the qualitative features such as singularity resolution because of a transition surface and the subsequent emergence of anti-trapped region are robust, \emph{quantitative} predictions for space-time geometry emerging from this dynamics can not be trusted because those numbers have no `gauge-invariant' meaning.  The origin of this $L_{o}$ dependence can be traced back  directly to the choice of $\dc$ in (\ref{munot}).  Since the effective equations contain (trigonometric) functions of $c\dc$ and, as we saw in section \ref{s2}, it is only $c/L_{o}$ that is invariant under the rescaling $L_{o} \to \alpha L_{o}$ of the fiducial cell, constancy of $\dc$ implies that the solutions to effective equations can carry an $L_{o}$ dependence. 

Our systematic numerical investigation revealed another limitation that is more subtle but conceptually equally important. If $\delta_{b}, \delta_{c}$ are assumed to be constant, then quantum effects can become important even in low curvature regime.  For large black holes, the Kretschmann scalar $K$ at and near the black hole horizon is very small. Already for  $m = 10^{5}$, we have $K = 7.5\times 10^{-21}$ at the horizon. In this approach, the effective space-time metric agrees with the classical metric to an excellent approximation till the curvature grows to $K \sim 10^{-19}$ but then coefficients of the two metrics start deviating and by the time the curvature becomes $K \sim 10^{-18}$ they are quite different from one another. \\

\emph{Remark:}  The $\mu_{o}$-type approach is used also in \cite{cgp} where, however, a deparametrization is carried out using the phase space variable $c$ as the internal time and a quantum corrected  effective description is obtained using an eikonal approximation.
Qualitatively the results are similar to those of the effective description summarized above. In particular $p_{c}$ undergoes a bounce. In our terminology, the parameters $\db, \dc$ are set equal to a numerical value as in \cite{ab}. Therefore the detailed predictions are again sensitive to the choice of $L_{o}$. 
 
\subsubsection{$\bar\mu$-type approaches}
\label{s4.D.2}

In FLRW models, limitations of the $\mu_{o}$ scheme were overcome through the so-called $\bar\mu$ scheme \cite{aps3}.  Soon thereafter, the technical ideas behind the $\bar\mu$ scheme were applied to the Schwarzschild interior in Refs. \cite{bv,dc} in the framework of effective theories. The key difference from \cite{ab,lm} is that $\db,\dc$ are no longer constants on the phase space: they are now phase space-functions
 \be \label{mubar}
 (\db)^{2} =  \f{\Delta}{p_{c}} \quad {\rm and} \quad   L_{o}^{2} (\delta_{c})^{2} = p_{c}\, \Delta\, \f{L_{o}^{2}}{p_{b}^{2}}.\ee
 %
(Note that,  as needed, $\db, L_{o}\dc$ are invariant under the rescaling $L_{0} \to \alpha L_{o}$ of the fiducial cell because $p_{b}/L_{o}$ is invariant.) Since $\dc$ depends on both $p_{c}$ and $p_{b}$, now the dynamical equations in the $b, p_{b}$  sector no longer decouple from those in the $c,p_{c}$ sector.  As a consequence, it has not been possible to write down analytical solutions; all explorations of the Schwarzschild interior in this approach have therefore been numerical \cite{bv,dc,dc2,js}.
 
In FLRW models, while physical results can depend on choices of fiducial structures in the $\mu_{o}$ scheme,  this is no longer the case in the $\bar\mu$ scheme \cite{aps3}. Similarly, while quantum corrections can become important in low curvature regions in the $\mu_{o}$ scheme, this does not occur in the $\bar\mu$ scheme. Therefore, initially it was hoped that these limitations of \cite{ab,lm} would be absent in \cite{bv,dc}. This expectation was borne out in part: The dependence on fiducial structures is indeed removed. However, the effective theory still has the second problem: there can be large quantum corrections in the low curvature region near the black hole horizon.  This can be most readily seen as follows. In the classical theory, since the space-like $\tau={\rm const}$ surfaces become null in the limit $\tau \to 2m$,  and in the Kantowski-Sachs model,  the expansion and anisotropic shears are calculated using the \emph{unit} time-like normal to the $\tau={\rm const}$  surfaces. Therefore  they diverge in the limit $\tau \to 0$ even though curvature at the horizon is very small for large black holes. The $\bar\mu$-type scheme does not distinguish between these harmless divergences and genuine singularities where space-time curvature diverges. $\bar\mu$-type schemes trigger large quantum corrections that make expansion and shears finite \cite{js} even at the horizon. Consequently, there are large departures from the classical theory very near the horizon. In terms of space-time metric, our numerical evaluations show  the same phenomenon in the dynamical behavior of $b, p_{b}$, both of which vanish classically at the horizon.  In the phase space, the classical and the effective trajectories differ significantly from each other, but only when one is \emph{very} close to the horizon.\\

\emph{Remark}: Had we been interested in Kantowski-Sachs cosmologies rather than the Schwarzschild interior, we would introduce matter fields in the form of a perfect fluid. Then, in classical general relativity the horizon is replaced by a (pancake-type) curvature singularity.  The large quantum corrections unleashed by the $\bar\mu$ scheme would then be physically appropriate.\\

If one moves away slightly from the horizon but remains in the low curvature region, the effective trajectories of the scheme proposed in Ref. \cite{bv} agree with the classical trajectories.  When the curvature reaches Planck scale,  $p_{c}$ reaches a minimum and bounces, giving rise to a transition surface $\T$. (Interestingly, $\T$ emerges somewhat before it does in the scheme proposed in this paper, i.e., when the curvature is smaller.)  As in section \ref{s2}, to the past of $\T$ we have a trapped region and to the future an anti-trapped region in which $p_{c}$ increases. \emph{However, the later evolution is qualitatively different from that in sections \ref{s4.B} and \ref{s4.C}} (see Fig. \ref{fig:3}). Now, the untrapped region is very small because  $p_{c}$ undergoes \emph{another} bounce and starts decreasing. Consequently, there is a new transition surface to the future of which we now have a trapped region. In this region $p_{c}$ decreases. But  the model becomes \emph{self-inconsistent} because very soon $p_{c}$ has decreased  so much that the round 2-spheres have an area smaller than $\Delta$ whence the required plaquette $\Box(\theta,\phi)$ cannot be fitted on the 2-sphere. That is, the requirements of the $\bar\mu$-scheme can no longer be implemented, whence its dynamical predictions cease to be meaningful. If one nonetheless continues the evolution as a mathematical exercise, one finds that the space-time geometry asymptotically approaches that of the Nariai {type} solution {\cite{bv,djs}}. However, conceptually this last prediction is not meaningful because, strictly speaking, the $\bar\mu$ scheme stops being applicable long before this stage is reached.  Physically, the scheme is not useful to explore the Schwarzschild interior because it sends the dynamical trajectory to phase space points where it ceases to be applicable. In this sense it fails by its own criteria.\\

\emph{Remark:} Whereas Ref. \cite{bv} considers vacuum Kantowski-Sachs space-times that are directly relevant for the analysis of Schwarzschild interior, Refs. \cite{dc,js} introduce matter sources. Similarly, in Ref. \cite{bkd,djs}, Kantowski-Sachs models with cosmological constant were studied along with a parallel treatment of locally rotationally symmetric (LRS)  Bianchi-III  spacetimes. These analyses used the $\bar\mu$ prescription but their results are not directly relevant to the Schwarzschild interior studied here. Indeed, the focus there was to probe issues related to the Kantowski-Sachs cosmology such as whether the singularity is resolved for general matter fields, whether it is possible to single out preferred quantization schemes in these cosmologies, e.g.,  by requiring that expansion and shears remain bounded, and other issues that had not been studied in LQC.

\subsubsection{Generalization of the $\mu_{o}$-type approach allowing mass dependence}
\label{s4.D.3}

To improve upon this situation, one can make a different choice of the quantum parameters $\db, \dc$ \cite{cs,oss}. 
{Ref. \cite{cs} modified the earlier $\mu_o$-type prescription using dimensional considerations and made it free of choice of fiducial  structures, while choices made in Ref. \cite{oss} were catered to obtain a symmetric bounce by phenomenologically modifying the scheme in Ref. \cite{cs}.  These choices can be viewed as lying} `in between' the $\mu_{o}$ and $\bar\mu$ schemes because they ask that $\db,\dc$ be phase space functions that are constant along any given dynamical trajectory,  but allow them to vary from one dynamical trajectory to another. Then, as we discussed in section \ref{s2.B}, the effective equations in the $b,\db$ sector decouple from those in the $c,\dc$ sector and the solutions are given by (\ref{eq:c}) -- (\ref{eq:pb}). 

Recall that under the rescaling $L_{o}\to \alpha L_{o}$, the connection component $b$ is invariant but $c$ changes via $c \to c/\alpha$. Since $b$ and $c$ enter the effective equations only via (trigonometric) functions of  $b\db$ and $c\dc$, to ensure cell independence of their solutions one needs to specify $\db$ and $L_{o}\dc$ in a way that they do not make reference to fiducial structures. Since $m := { \sin(\dc c) p_c}/{(\dc L_o \gamma)} $ is a constant of motion (see (\ref{m})),  using dimensional considerations $\db,\dc$ were set to
\be \label{cs}
(\db)^{2} = \f{\Delta}{(2m)^{2}}  \quad  {\rm and}\quad  L_{o}^{2} (\dc)^{2} = \Delta. \ee
Thus, as in the $\mu_{o}$ scheme, $\dc$ is constant on the entire phase space, but $\delta_{b}$ now depends on $m$. It thus varies from one dynamical trajectory to another.  Although the area gap $\Delta$ features in the expressions, the ansatz is motivated by phenomenological rather than fundamental considerations because  one does not specify how the plaquettes $\Box(\theta,\phi),\, \Box(x,\theta),\, \Box(x,\phi)$ enclosing area $\Delta$ are to be chosen. Rather,  the prescription  (\ref{cs})  was made because it is the simplest one that is dimensionally consistent and meets the `cell-independence' requirement. 

Physical predictions of the effective dynamics resulting from this modified $\mu_{o}$ scheme have several attractive features \cite{cs}. First, by design they are all insensitive to the choice of $L_{o}$. Second, it again follows from (\ref{eq:pc}) that $p_{c}$ has one and only one minimum.  In the space-time picture, this again implies that the extended effective space-time is divided into a trapped and an anti-trapped region, separated by the transition surface $\T$ (which corresponds to the absolute minimum of $p_{c}$).  Third, the anti-trapped region has a future boundary that corresponds to a white hole type horizon. Finally, unlike in the $\bar\mu$ scheme, the expansion and shear grow unboundedly near the black hole horizon; for large black holes, the space-time geometry  near horizons is well approximated by general relativity. Thus the space-time picture is qualitatively similar to that in sections \ref{s3}, \ref{s4.B} and \ref{s4.C}. 

However, there are two major differences.  First, for large black holes, while  the trapping surface $\T$ always emerges in the Planck regime in our approach,  in this generalized $\mu_{o}$ scheme, it  can emerge in low curvature regime. In fact the curvature at $\T$ goes to zero in the limit $m\to \infty$. Thus, for astrophysical black holes very large quantum effects  arise in low curvature regions. This feature can be qualitatively understood as follows.  Eq. (\ref{eq:pc}) implies that at the transition surface  $\T$, we have $p_{c}|_{\T} =  m (\gamma\, L_{o}\dc)$.  Since in this approach $L_{o}\dc = \sqrt{\Delta}$,  we have: $p_{c}|_{\T} = (\gamma\sqrt\Delta)\, m$. Now,  in \emph{classical general relativity} the Kretschmann scalar{}%
\footnote{In the effective theory, the expression of the Kretschmann scalar is much more complicated. However, for large $m$, the effective trajectory is well approximated by the classical one until it reaches close to the $p_{c}$-bounce.  (For example, even for a rather low value of mass, $m=10^{5}$, the transition surface emerges at $T=-7.1$ while the two trajectories are indistinguishable between $T=0$ and $T=-6$.) Therefore the classical expression of $K$ provides a very good approximation to the actual value. } 
is given by $K_{\rm cl} = 48 m^{2}/p_{c}^{3}$. Therefore, at the bounce surface $K_{\rm cl} = 48/(\gamma^{3} \Delta^{3/2}\, m)$; it \emph{decreases} as $1/m$. This dependence is borne out in numerical simulations. \\

\emph{Remark:} The requirements of cell-dependence and dimensional consistency criteria do restrict the choice of $\db,\dc$, but they still leave considerable freedom because $m^{2}/\Delta$ is dimension-free. Our proposal for $\db,\dc$ of section \ref{s4.A} also satisfies these criteria. However, now the transition surface $\T$ always emerges in the Planck regime, and, furthermore, curvature scalars have an absolute, mass independent upper bounds.  These features can also be understood using the same general considerations. As noted above, if $\db,\dc$ are Dirac observables, then at the transition surface $p_{c}$ is given by $p_{c}|_{\T} =  m (\gamma\, L_{o}\dc)$. In our choice (\ref{db-dc}), we have  $L_{o} \dc = C m^{-1/3}\,\Delta^{{2/3}}$ where $C$ is a dimensionless constant.  Therefore it follows that the classical Kretschmann scalar is now a Planck scale, mass independent constant: $K_{\rm cl} = (48/\gamma^{3} C^{3} \Delta)$. While  the Kretschmann scalar of the effective metric has a much more complicated form, as we showed in section \ref{s4.C},  its leading term is also mass independent  and of Planck scale for the macroscopic black holes we are interested in. \\

The second major difference between this generalized $\mu_{o}$ scheme and the one introduced in this paper is the following. In our approach the radius of the white hole-type horizon is the same as that of the initial black hole horizon in the large $m$ limit.  As we will see in section \ref{s5}, this implies that the ADM mass in asymptotic region III associated with the white hole horizon agrees with that in the asymptotic region I associated with the initial black hole horizon: 
\be m_{\rm WH} = m_{\rm BH}\,  \Big(1+{\cal O} \big( (\f{\lp}{m})^{\f{2}{3}} \, \ln (\f{m}{\lp}) \big) \Big). \ee
In the generalized $\mu_{o}$ scheme, on the other hand, there is a tremendous mass amplification, approximately given by \cite{cs}
\be  \label{inflation} m_{\rm WH}\,\,  \approx \,\, (m_{\rm BH}) \,\Big( \f{m_{\rm BH}}{\lp}\Big)^{3}. \ee
 For a solar mass black hole this would be an increase by a factor $\sim10^{114}$! The physical origin of this huge increase has remained unclear.
 
Finally, Ref.  \cite{oss} studied the possibility of removing the mass amplification within the broad idea of using a generalized $\mu_{o}$-scheme but modifying the  ansatz (\ref{cs})  to
\be  \label{oss}
(\db)^{2}  = \alpha^{2} \f{\Delta}{(2m)^{2}} \quad  {\rm and}  \quad L_{o}^{2} (\dc)^{2} =  \beta^{2} \,\Delta, \ee
where $\alpha,\beta$ are dimensionless constants. Again, the approach is phenomenological in the sense that there is no prescription to choose the plaquettes that are to enclose the area $\Delta$. Rather, the idea was to first stipulate conditions on $\alpha$ and $\beta$ to reduce the freedom to a single constant and fix that remaining freedom by imposing the requirement that the mass amplification factor be $1$. Three possibilities were explored: (i) $\beta=1$;\, (ii) $\alpha = 1$\, and (iii) $\alpha\beta =1$. Because the final goal of arriving at the mass amplification factor of $1$ is also realized in our approach in the large $m$ limit,  there is some similarity between the two. However, there are also differences. At the conceptual level,  our choice (\ref{db-dc}) of Dirac observables $\db,\dc$ was arrived at by specifying the plaquettes. At a practical level, none of the three choices of \cite{oss} is compatible with our choice (\ref{db-dc}).  For example, with choices (i) and (iii) the  forms of $\db,\dc$ are not known analytically even for the large $m$, while in our approach they are given simply by (\ref{db-dc}). For choice (ii), the asymptotic forms for large $m$ are given in \cite{oss}, and they imply $p_{c} |_{\T} \approx (\gamma\Delta^{2})/ m^{2}$ at the transition surface. Hence now the classical expression of the Kretschmann scalar $K_{\rm cl} = 48m^{2}/p_{c}^{3} =C m^{8}/\Delta^{4}$ (where $C$ is a dimensionless constant)  grows unbounded with $m$. In our approach, it has a mass independent upper bound.

\section{Quantum extended Kruskal space-time} 
\label{s5}
This section is divided into two parts. In the first we introduce a new approach to obtain the quantum corrected effective metric in the `exterior region' between the horizon and infinity using  LQG techniques as in section \ref{s2}. In the second we show that the effective metric in the `interior' and the `exterior' regions match seamlessly and investigate properties of the resulting quantum extension of Kruskal space-time.

\subsection{The Schwarzschild exterior} 
\label{s5.A}

\subsubsection{Phase space for the exterior region}
\label{s5.A1}

As we saw in section \ref{s2},  a finite dimensional phase space can be constructed for the interior region using the fact that it is foliated by spatially homogeneous space-like 3-manifolds $\Sigma$. Since phase spaces are normally constructed using Cauchy surfaces and since none in the exterior region are homogeneous, Hamiltonian descriptions of the exterior have been qualitatively different. On the one hand, they are much more complicated because the inhomogeneity of the spatial metric makes these standard phase spaces infinite dimensional. On the other hand since the exterior is static, discussion of dynamics is somewhat vacuous. 

Our new observation that changes this status quo is rather simple. While the exterior cannot be foliated by space-like homogeneous surfaces, it \emph{is} foliated by \emph{time-like} homogeneous surfaces ($r={\rm const}$ in the standard Schwarzschild coordinates) whose isometry group is again ${\rm R} \times {\rm SO(3)}$.  Therefore the phase space based on these slices is again finite dimensional and there is now non-trivial `dynamics' as one `evolves' from one time-like homogeneous surface to another in the radial direction. While this is somewhat counter-intuitive at first because this `evolution' is in a space-like direction, there is nothing unusual about the setup from the Hamiltonian perspective  even for full general relativity: One again has `constraint equations' on the canonical variables, and `dynamics' is again generated by a Hamiltonian constraint.  Indeed, such `evolutions' in spatial directions are already used extensively in LQG in the context of Euclidean/Riemannian frameworks. Of course, considerable work is needed to extend the LQG framework to cover this situation and it is far from being clear that all potential problems can be handled satisfactorily.%
\footnote{Several months after the first version of this paper was posted on the arXiv, we became aware that this basic idea was already put forward by Liu and Noui in 2017  \cite{hlkn}.  Note, however, that in this paper we have restricted ourselves to the homogeneous context, where some of the key difficulties (associated with cylindrical-consistency in presence of non-compact internal groups)  are bypassed. Note also that our prescription to select $\delta_{b}, \delta_{c}$ (spelled out in section \ref{s4.A}) requires quantum geometry considerations \emph{only} at the transition surface $\T$ which lies in the `interior region' where homogeneous slices are space-like.}

In the Kruskal space-time now under consideration,  not only will the  `dynamics' again be generated by a Hamiltonian constraint, but the `evolution equations' will again be ODEs as in section \ref{s2}. The only difference from the situation in section \ref{s2} is that now the intrinsic metric $q_{ab}$ has signature -,+,+ (rather than +,+,+) and therefore the internal space for the gravitational connection and triads also has signature -,+,+ (rather than +,+,+) . This means that the gauge group of internal rotations for the gravitational connection is now ${\rm SU(1,1)}$ (rather than ${\rm SU(2)}$).  While a convenient basis in the Lie algebra of ${\rm SU(2)}$ is given by $\tau_{i}$ used in Eqs (\ref{connection}) and (\ref{triad}),  for ${\rm SU(1,1)}$  it is given by $\t\tau_{i}$, related to $\tau_{i}$ via:
\be \label{conversion}
\t\tau_{1} = i \tau_{1}, \quad \t\tau_{2} = i \tau_{2}, \quad \t\tau_{3} =  \tau_{3}, \ee

Keeping this difference in mind, we can simply follow the procedure used in section \ref{s2} step by step. Let us therefore consider a 3-manifold $\t\Sigma$  again with topology $\mathbb{R}\times \mathbb{S}^{2}$ and introduce on it a fiducial metric
\be  \label{fiducialt}  \mathring{\tilde{q}}_{ab} \d x^{a} \d x^{b}  =  -\d x^{2} +  r_o^2(\d \theta^2 + \sin^2 \d \phi^2), \ee
where, again,  $x \in (-\infty,\infty)$, \, $\theta$ and $\phi$ are 2-sphere coordinates and $r_{o}$ is a constant. (Note that $x$ is now a time-like coordinate, $\partial/\partial x$ being the time translation Killing field in the exterior region.) Then, thanks to the underlying homogeneity we can solve the spatial diffeomorphism constraint and perform a partial gauge fixing to satisfy the Gauss constraint. As a result,  the gravitational connection and the conjugate densitized triad can be expressed as in equations (\ref{connection}) and (\ref{triad}) simply by replacing $\tau_{i}$ by $\t\tau_{i}$ and using the relation (\ref{conversion}) between the two:
\be \label{connectiont}
A^i_a \, \t\tau_i \, \d x^a \, = \, \f{\t{c}}{L_{o}}  \, \tau_3 \, \d x +  i\t{b} \, \tau_2 \d \theta 
- i\t{b}\, \tau_1 \sin \theta \, \d \phi + \tau_3 \cos \theta \, \d \phi,
\ee
and
\be \label{triadt}
E^a_i \, \t\tau^i \partial_a \, =  \, \t{p}_c \, \tau_3 \, \sin \theta 
\, \partial_x + \f{i\t{p}_b}{ L_o} \, \tau_2 \, \sin \theta \, 
\partial_\theta -\f {i\t{p}_b}{ L_o} \, \tau_1 \,  \partial_\phi . 
\ee
Comparing these equations with (\ref{connection}) and (\ref{triad}) it is clear that  the phase space for the exterior region can be obtained simply by making the substitutions 
\be \label{substitutions} b \to i\t{b},   \, p_{b} \to i \t{p}_{b}; \qquad c\to \t{c},\, p_{c} \to \t{p}_{c}, \ee
in equations of section \ref{s2}.  In particular, the Poisson brackets  are now given by:
\be\label{pbst}
\{\t{c},\, \t{p}_c\} \, = \,2 G \gamma, \quad \{\t{b},\, \t{p}_b\} \, = \,  -G \gamma.
\ee

\subsubsection{Classical dynamics of the exterior region}
\label{s5.A2}
Let us now turn to dynamics. From space-time perspective, since the radial coordinate $\tau$ satisfies $\tau >2m$ in the exterior region, the Hamiltonian  dynamics simply `evolves' the geometry in radial directions filling out the exterior region $2m< \tau < \infty$, starting from the data at some $\tau_{0} >2m$. However, it is instructive to examine this evolution as a dynamical trajectory in phase space as a prelude to the investigation of the  effective dynamics.

We begin by writing the Hamiltonian constraint for the exterior region, obtained simply by using the substitutions (\ref{substitutions}) in (\ref{Hcl}):
\be\label{Hclt}
\t{H}_{\mathrm{cl}}[\t{N}_{\rm{cl}}] = - \f{1}{2 G \gamma}\left(2 \t{c} \, \t{p}_c + \left(-\t{b} + \f{\gamma^2}{\t{b}}\right) \t{p}_b  \right) .
\ee
As one would expect, evolution equations for connection and triad variables obtained using  (\ref{pbst}) and (\ref{Hclt}) are the same as those that result if one uses the substitutions (\ref{substitutions}) in (\ref{conf-dot}) and (\ref{mom-dot}). One can easily integrate these equations and use (\ref{Hclt}) to simplify the solutions. The result is:
\be \label{conft}
\t{b}(T_{\rm{cl}})=\pm \gamma\, \left(1- e^{-T_{\rm{cl}}}\right)^{1/2}, \quad \t{c}(T_{\rm{cl}}) =  \t{c}_o\,e^{-2 T_{\rm{cl}}},
\ee
and
\be \label{momentat}
\t{p}_b(T_{\rm{cl}})=\t{p}_b^{(o)}\, e^{T_{\rm{cl}}}\,\big(1- e^{-T_{\rm{cl}}} \big)^{1/2}, \qquad \t{p}_c(T_{\rm{cl}}) = \t{p}_c^{(o)}\,e^{2 T_{\rm{cl}}}, \ee
where $T_{\rm cl}$ is the affine parameter along the Hamiltonian vector field generated by (\ref{Hclt}).  The form of the solutions (\ref{conft}) and (\ref{momentat}) immediately implies that  $\t{c}\, \t{p}_c/(L_o \gamma)$ is a Dirac observable, i.e., it is a constant of motion.  As we will see,
in the space-time metric associated with any dynamical trajectory, it again coincides with  $m = GM$.

As in section \ref{s2.A},  we have fixed one of the integration constants  so that  the black hole horizon lies at $T_{\rm cl}=0$ in the space-time picture. The remaining integration constants $\t{c}_o$, $\t{p}_b^{(o)}$ and $\t{p}_c^{(o)}$ can also  be fixed as in section \ref{s2.A} to match the phase space variables with the corresponding space-time geometry in the Schwarzschild exterior: 
\be\label{metric-kruskalt}
\d \t{s}^2 =  - \Big(1-\frac{2 m}{\tau} \Big) \d x^2 + \Big(1-  \frac{2 m}{\tau} \Big)^{-1} \d \tau^2 + \tau^2 (\d \theta^2 + \sin^2\theta \d \phi^2). 
\ee
To set this correspondence, we  first note that for any choice of the radial coordinate $\tau$ and the associated `lapse-squared' $\t{N}^{2}_{\tau}$,  each point in the phase space defines a space-time metric admitting a foliation by  homogeneous time-like slices:
\be\label{metrict}
\t{g}_{ab} \d x^{a} \d x^{b}  \equiv \d \t{s}^2 =  - \f{\t{p}_b^2}{|\t{p}_c| L_o^2} \d x^2 - \t{N}_{\tau}^2 \d \tau^2+ |\t{p}_c| (\d \theta^2 + \sin^2\theta \d \phi^2).
\ee
(See Eq. (\ref{metric}). As we will see below, $\t{N}_{\tau}^{2}$ is negative, reflecting the fact that $\tau$ is a space-like rather than a time-like coordinate.)  Our solutions (\ref{conft}) and (\ref{momentat}) are written for a specific choice $T_{\rm cl}$ of the radial coordinate in the exterior. As in section \ref{s2.A}, the transformation relating $T_{\rm cl}$ to the radial coordinate $\tau$ in (\ref{metric-kruskalt}) is:\, $\tau := 2 m e^{T_{\rm{cl}}}$.  With this information at hand, we can now use (\ref{conft}) and (\ref{momentat})  in (\ref{metrict}) and set up the desired dictionary by comparing the resulting expression with (\ref{metric-kruskalt}).  Comparing the first and the last terms in these two expressions of the metric we obtain \, $|\t{p}_{c}| = \tau^{2}$ and $\t{p}_{b}^{2} = L_{o}^{2} (1- \f{2m}{\tau})\tau^{2}$. With this choice of $\tau$, the lapse takes the form $\t{N}_{\tau}^{2} = - (1-\f{2m}{\tau})^{-1}$. Therefore,  the remaining constants are given by:  {$\t{c}_{o}= \gamma L_{o}/4m; \,\, \t{p}_{b}^{(o)} = {2m} L_{0};\,\, \tilde p_{c}^{(o)} = 4m^{2}$.}  As in footnote \ref{fn5}, the more familiar Schwarzschild form is obtained by the obvious substitutions $x \to t$ and $\tau \to r$. 

Finally we note that at the horizon we have $\tau =2m$, whence as in the interior solution $T_{\rm cl} $ vanishes there and so do $\t{p}_{b}$ and $\t{b}$. Thus, in the phase space picture the horizon is characterized by the same conditions in both interior and exterior regions.  We will see in section \ref{s5.B1} that the matching is seemless both in the phase space and space-time pictures.

\subsubsection{Effective dynamics  of the exterior region.}
\label{s5.A3}

Let us now turn to effective dynamics on the phase space of section \ref{s5.A1} associated with the exterior region. We can now  just follow the analysis of section \ref{s2.B} step by step. Substitutions (\ref{substitutions}) imply that the effective Hamiltonian constraint is given by:
\be \label{H_efft} 
\t{H}_{\mathrm{eff}}[\t{N}] =  - \f{1}{2 G \gamma} \Big[2 \f{\sin (\delta_{\t{c}}\, \t{c})}{\delta_{\t{c}}}  \, |\t{p}_c|  + \Big(-\f{\sinh (\delta_{\t{b}}\, \t{b})}{\delta_{\t{b}}} + \frac{\gamma^2 \delta_{\t{b}}}{\sinh(\delta_{\t{b}}\, \t{b})} \Big) \, \t{p}_b  \Big] ~.  \ee
where $\delta_{\t{b}} = \delta_{b},\,\, \delta_{\t{c}} = \delta_{c}$ continue to be given by 
(\ref{db-dc}). Thus, the same principle determines these quantum parameters both in the interior and the exterior. Note that the expression on the right side now involves trigonometric functions of $\delta_{\t{c}}\, \t{c}$ but hyperbolic functions of $\delta_{\t{b}}\, \t{b}$, reflecting the fact that the $x$-direction is now time-like rather than space-like while $\theta,\phi$-directions continue to be space-like. One can obtain the equations of motion using this Hamiltonian constraint and Poisson brackets (\ref{pbst}) and find their solutions.

As one would expect, the solutions are the same as those resulting from substitutions (\ref{substitutions}) in the interior solution (\ref{eq:c}) -- (\ref{eq:pb}):
\ba \label{eq:ct}
\tan \Big(\f{\delta_{\t{c}}\, \t{c}(T)}{2} \Big)&=&  \mp \f{\gamma L_o \delta_{\t{c}}}{8 m} e^{-2 T},\\
\label{eq:pct} \t{p}_c(T) &=& 4 m^2 \Big(e^{2 T} + \f{\gamma^2 L_o^2 \delta_{\t{c}}^2}{64 m^2} e^{-2 T}\Big) ,
\ea  
\be \label{eq:bt}
\cosh \big(\delta_{\t{b} }\,\t{b}(T)\big) = \t{b}_o \tanh\left(\f{1}{2}\Big(\t{b}_o T + 2 \tanh^{-1}\big(\frac{1}{\t{b}_o}\big)\Big)\right),
\ee
where%
\be
\t{b}_o = (1 + \gamma^2 \delta_{\t{b}}^2)^{1/2} ,
\ee
and,
\be\label{eq:pbt}
\t{p}_b(T) = - 2 \, \f{\sin (\delta_{\t{c}}\, \t{c}(T))}{\delta_{\t{c}}} \f{\sinh (\delta_{\t{b}}\, \t{b}(T))}{\delta_{\t{b}}} \f{|\t{p}_c(T)|}{\gamma^{2 }-\f{\sinh^2(\delta_{\t{b}}\, \t{b}(T))}{\delta_{\t{b}}^2} },
\ee
where $T$ is now the affine parameter along the Hamiltonian vector field generated by $\t{H}_{\rm eff} [\t{N}]$. Note that, just as the $b,p_{b}$ and $c,p_{c}$ sectors decouple dynamically in the Schwarzschild interior, the   $\t{b}, \t{p}_{b}$ and $\t{c},\t{p}_{c}$ sectors also decouple in the exterior.  The form of solutions in the $\t{c}, \t{p}_{c}$ sector is the same as that in the $c,p_{c}$ sector, while in the $\t{b}, \t{p}_{b}$ sector, up to some changes in signs, trigonometric functions (such as $\sin (\delta_{b} b)$) are replaced by hyperbolic functions (such as $\sinh(\delta_{\t{b}} \t{b})$).  Because of the agreement of dynamics in the $\t{c}, \t{p}_{c}$ and $c, p_{c}$ sectors,\, $\sin(\delta_{\t{c}} \t{c}) \t{p}_c/(\delta_{\t{c}}\, L_o \gamma)$  continues to be a Dirac observable which in the classical regime has the interpretation of $m= GM$ of the black hole. Since our effective theory agrees with the classical theory in the low curvature region near the black hole horizon and since  Dirac  observables are constants of motion, as in section \ref{s2.B}, we will again denote it by $m$. 

Finally, one can pass from the phase space to the space-time description following the same procedure as in the classical theory, sketched in section \ref{s5.A2}.  The space-time metric $\t{g}_{ab}$ is of the form  (\ref{metrict}) 
\be 
\t{g}_{ab} \d x^{a} \d x^{b} =  - \f{\t{p}_b^2}{|\t{p}_c| L_o^2} \d x^2 + \f{\gamma^{2} |\t{p}_{c}|\, \delta_{\t{b}}^{2}}{\sinh^{2} (\delta_{\t{b}}\t{b})} \d T^2+ |\t{p}_c| (\d \theta^2 + \sin^2\theta \d \phi^2).
\ee
since  $\t{N}^{2}$ now  has the form
\be \label{Nt}
 \t{N}^{2} = - \, \f{\gamma^{2} |\t{p}_{c}|\, \delta_{\t{b}}^{2}}{\sinh^{2} (\delta_{\t{b}}\t{b})}.\ee
Explicit expressions of $\t{p}_{c},\, \t{p}_{b}$ are given by (\ref{eq:pct}) and (\ref{eq:pbt}),

\subsection{Properties of the quantum extension of the Kruskal space-time}
\label{s5.B}

From the Hamiltonian perspective, we have two distinct phase spaces, one spanned by $b,p_{b}; \, c,p_{c}$ and another by $\t{b}, \t{p}_{b};\, \t{c}, \t{p}_{c}$, with Poisson brackets given by (\ref{pbs}) and (\ref{pbst}). Dynamics is generated by distinct Hamiltonian constraints -- (\ref{Hcl})   and (\ref{Hclt}) in the classical theory and  (\ref{H_eff}) and  (\ref{H_efft}) in the effective theory. 
Nonetheless,  as we show in section \ref{s5.B1} the space-time geometries defined in the exterior and interior regions match smoothly across horizons. In section \ref{s5.B2} we investigate properties of the resulting quantum extension of the Kruskal space-time.

\nopagebreak[3]\begin{figure} [h] 
\includegraphics[width=0.8\textwidth]{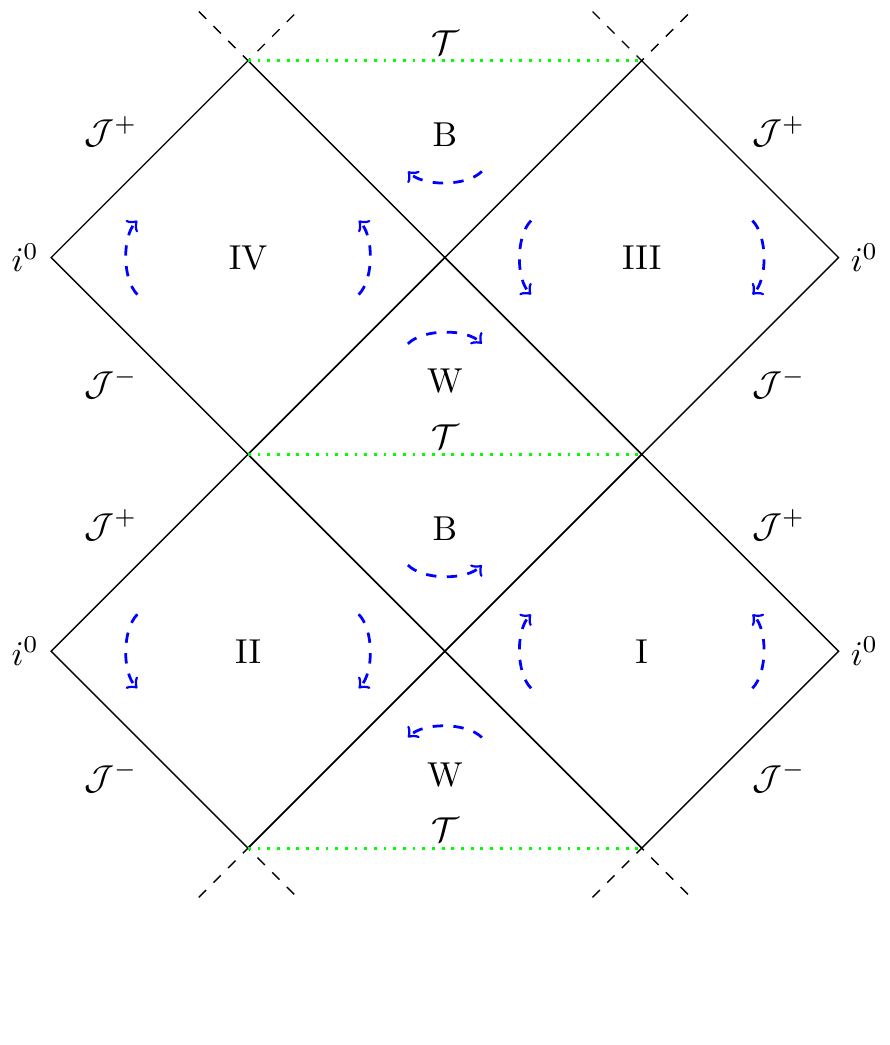}  
\caption{\footnotesize{The Penrose digram of the extended Kruskal space-time.  In sections \ref{s3} and \ref{s4}   we began with the region B in the central diamond corresponding to the standard  Schwarzschild trapped region. In the classical theory space-time ends at  a horizontal line due to curvature singularity. In the effective geometry, the singularity is replaced by a transition surface $\T$ where $\dot{p}_c$ vanishes. The extended region has an anti-trapped region labeled W. Thus quantum geometry provides a region bounded to the past by a black hole type horizon and to the future by a white hole type region.  Effective metric for asymptotic regions I, II, III and IV is introduced in section \ref{s5.A3}.  Section \ref{s5.B1} shows that it joins on smoothly to the geometry in the trapped and anti-trapped regions B and W.  As indicated by dashed lines, the same procedure continues the space-time to new trapped and anti-trapped and asymptotic regions to the past and future. Arrows denote trajectories of the translational (or static) Killing field $X^{a}\partial_{a} = \partial_{x}$ in various regions.} }
\label{fig:4}
\end{figure}

\subsubsection{Matching of interior and exterior geometries}
\label{s5.B1}

Let us begin with the classical theory where the situation is straightforward. Dynamical trajectories in the interior phase space correspond to time $T_{\rm cl} < 0$ while those in the exterior phase space correspond to $T_{\rm cl}>0$. In the space-time interpretation, the $T_{\rm cl}=0$ surface is excluded in both descriptions since it is null. However, one can regard it as a limit of space-like $T_{\rm cl}={\rm const} <0$ surfaces in the interior region and time-like $T_{\rm cl}={\rm const} >0$ surfaces in the exterior region and in both regions the limit represents a black hole type horizon. Therefore we can ask whether the geometry is smooth across this horizon. The triad variables are given, respectively, by
\be T_{\rm cl} <0 : ~~
p_b(T_{\rm{cl}})=p_b^{(o)}\, e^{T_{\rm{cl}}}\,\big(e^{-T_{\rm{cl}}} - 1\big)^{1/2}, \qquad p_c(T_{\rm{cl}}) = p_c^{(o)}\,e^{2 T_{\rm{cl}}} \ee 
and 
\be   T_{\rm cl} >0 : ~~
\t{p}_b(T_{\rm{cl}})=\t{p}_b^{(o)}\, e^{T_{\rm{cl}}}\,\big(1- e^{-T_{\rm{cl}}} \big)^{1/2}, \qquad \t{p}_c(T_{\rm{cl}}) = \t{p}_c^{(o)}\,e^{2 T_{\rm{cl}}}, \ee
where $p_{b}^{o} = \t{p}_{b}^{ o} = { 2m} L_{0} $  and $p_{c}^{o} = \t{p}_{c}^{ o} = 4m^{2}$. Therefore, it is evident that the triad variables are smooth across the boundary $T_{\rm cl}=0$. Thus, the dynamical trajectory in the Schwarzschild interior can be joined smoothly with that in the exterior provided, of course, they both correspond to the same mass.

As one would expect, the situation for the space-time metric is a bit more complicated simply because the  constant $T_{\rm cl}$ surfaces are space-like for $T_{\rm cl} <0$,  time-like  for  $T_{\rm cl} >0$ and null at $T_{\rm cl} =0$. So, we have:
\ba  
  \hskip-0.0cm T_{\rm cl} <0 &:&\, \,g_{ab}   \d x^{a} \d x^{b} = -  \f{4m^{2} e^{2 T_{\rm cl}}}{(e^{-T_{\rm cl}} -1)} \d T_{\rm cl}^{2} + (e^{-T_{\rm cl}} -1) \d x^{2} + 4m^{2} e^{2T_{\rm cl}}
\big(\d\theta^{2} + \sin^{2}\theta \d \phi^{2} ) \nonumber\\
\hskip-0.0cm  T_{\rm cl} >0 &:&\,\,  \t{g}_{ab} \d x^{a} \d x^{b} = -  (1- e^{-T_{\rm cl}} ) \d x^{2} +
 \f{4m^{2} e^{2 T_{\rm cl}}}{(1- e^{-T_{\rm cl}} )} \d T_{\rm cl}^{2} +4m^{2} e^{2T_{\rm cl}}
\big(\d\theta^{2} + \sin^{2}\theta \d \phi^{2} ), \nonumber\\
\ea
where we have used the explicit form of the lapse  $N_{\rm cl}^{2}$ and $\t{N}_{\rm cl}^{2}$ in the interior and the exterior regions (see  Eq. (\ref{Ncl})) .

In both regions the 4-metric  is ill-defined in the limit $T_{\rm cl} \to 0$. However, this is the standard coordinate singularity; space-time geometry is in fact smooth there. As is well-known, because the 2-sphere metric is smooth and non-degenerate at $T_{\rm cl} =0$ and the determinant $g_{xx}\,g_{T_{\rm cl}T_{\rm cl}} $ of the 2-metric in the $x$-$T_{\rm cl}$ plane  is smooth and \emph{non-vanishing} there, one can introduce the standard Eddington-Finkelstein coordinates to show explicitly that the 4-metric is manifestly smooth across the horizon $T_{\rm cl} =0$. This statement in terms of space-time geometry is trivial. For the effective theory  discussed below, the important observation is that smoothness of space-time geometry across the horizon is guaranteed by the following properties of the dynamical trajectories in the phase space: at $T_{\rm cl } =0,$ \\
 (1) the pairs $p_{b}, \t{p}_{b}$ and $p_{c}, \t{p}_{c}$ admit smooth limits;\\ 
(2)  $N^{2}_{\rm cl } = - \t{N}^{2}_{\rm cl}$, reflecting the fact that while $N^{2}_{\rm cl}$ refers to time-evolution in the interior region, $\t{N}^{2}_{\rm cl}$ refers to `radial-evolution' in the exterior region; and,\\ 
(3) $(N^{2}_{\rm cl}p_{b}^{2})/(p_{c}L_{o}^{2})$ in the interior and its counterpart $(-\t{N}^{2}_{\rm cl}\t{p}_{b}^{2})/(|\t{p}_{c}| L_{o}^{2})$ in the exterior are smooth, non-zero  and equal.  (Each of these quantities is the determinant of the 2-metric in the $x$-$T_{\rm cl}$ plane).\\

We will now show that these properties continue to hold also in the effective description. The expressions of triads along dynamical trajectories in the phase space are:

\ba T <0 : ~~ p_c(T) &=& 4 m^2 \Big(e^{2 T} + \f{\gamma^2 L_o^2 \dc^2}{64 m^2} e^{-2 T}\Big) , \nonumber\\
p_b(T) &=& - 2 m L_{o} \, \f{\sin (\db\, b(T))}{\db} \f{1}{\f{\sin^2(\db\, b(T))}{\db^2} + \gamma^2}\, ;
\ea
 and, 
\ba T >0: ~~ 
 \t{p}_c(T) &=& 4 m^2 \Big(e^{2 T} + \f{\gamma^2 L_o^2 \delta_{\t{c}}^2}{64 m^2} e^{-2 T}\Big) ,\nonumber\\ 
\t{p}_b(T) &=& - 2 m L_{o}\,  \f{\sinh (\delta_{\t{b}}\, \t{b}(T))}{\delta_{\t{b}}} \f{1}{\gamma^{2 }-\f{\sinh^2(\delta_{\t{b}}\, \t{b}(T))}{\delta_{\t{b}}^2} },
\ea
where we have simplified the expressions of $p_{b}(T)$ and $\t{p}_{b}(T)$ using  the form of the Dirac observable $m= (\sin \delta_{c}c)p_{c}/(\delta_{c} L_{o}\gamma) = (\sin \delta_{\t{c}}\t{c}) \t{p}_{c}/(\delta_{\t{c}} L_{o}\gamma)$.
Note that for matching the exterior and interior geometries,  we are interested in  pairs of trajectories in the interior and exterior phase space with the same value of $m$.  On these pairs, at $T =0$ we have: $p_{c} = \t{p}_{c} = 4m^{2}\Big(1 + \f{\gamma^2 L_o^2 \delta_{\t{c}}^2}{64 m^2} \Big)$ and $p_{b}/L_{o} = \t{p}_{b}/L_{o} = 0$. Thus, the values of the triad variables match at $T=0$ and by Taylor expanding them one can check that the matching is smooth. Therefore, condition (1) above is satisfied. Eqs (\ref{N}) and (\ref{Nt}) imply that condition (2) is also satisfied. Finally,  at $T=0$ we have:  $(N^{2} p_{b}^{2})/(p_{c}L_{o}^{2}) = (-\t{N}^{2}\t{p}_{b}^{2})/(|\t{p}_{c}| L_{o}^{2}) = 4m^{2}$ (which, incidentally is exactly the same as in the classical theory). Thus, condition (3) is also satisfied. \\

Let us conclude with a summary of the situation in the effective theory.  We have separate, 4-dimensional phase spaces describing the exterior and interior Schwarzschild geometries, depicted  in Fig \ref{fig:4} by  region I and the black hole region B attached to it. They are coordinatized, respectively, by pairs  $\t{b},\t{p}_{b}; \, \t{c}, \t{p}_{c}$  with Poisson brackets  (\ref{pbst}), and pairs $b,p_{b}; \, c, p_{c}$, with Poisson brackets (\ref{pbs}). Dynamics is governed by the Hamiltonian constraints (\ref{H_efft}) and  (\ref{H_eff}) respectively.  We identified a Dirac observable $m$ in each phase space. The parameter $T$ along trajectories $\t{b}(T), \t{p}_{b}(T);\, \t{c}(T), \t{p}_{c} (T)$ in the exterior phase space has positive values and along trajectories ${b}(T), {p}_{b}(T);\, {c}(T), {p}_{c} (T)$ in the interior phase space takes negative values. Trajectories labeled by the same value of  the Dirac observable $m$ can be joined smoothly at $T=0$. In the space-time language, the (limiting) point  $T=0$ along each trajectory represents the black hole horizon. The coefficients in the space-time metric become singular in the Schwarzschild-like coordinates $(T,x)$, just as they do in the classical theory. However, the effective metric is smooth across the horizon. The metric coefficients are such that the determinant of the effective 4-metric remains smooth and non-zero across $T=0$ whence, as in the classical theory, one can introduce new Eddington-Finkelstein type coordinates to show that the effective geometry is \emph{manifestly}  smooth in the entire region ${\rm I} \cup {\rm B} \cup{\rm W}$ of Fig \ref{fig:4}, which encompasses the asymptotic region I as well as the black and white hole regions B and W that are joined at the transition surface $\T$. For the macroscopic black holes considered in this paper, the asymptotic region I is  `tame' as in the classical theory. However, the interior region ${\rm B} \cup{\rm W}$ includes Planck scale curvature where quantum geometry effects resolve the singularity.

\subsubsection{Properties of the quantum extension}
\label{s5.B2}

The procedure introduced in section \ref{s5.A} can be used again at the left boundary of the black hole region B to extend the effective space-time to  the asymptotic region II. Similarly, the procedure can be used at the future boundaries of the white hole region W to join the space-time smoothly to asymptotic regions III and IV. These regions have future boundaries representing black hole type horizons and so we can again repeat the procedure and extend the space-time to future. Thus, the procedure provides a quantum extension of the Kruskal space-time where the effective space-time metric is everywhere smooth and curvature invariants are uniformly bounded. The full extension has infinitely many asymptotic, trapped and anti-trapped regions. This structure is shown in Fig. \ref{fig:4}.\smallskip

We will now discuss salient features of this quantum extension of Kruskal space-time.\smallskip

$\bullet$ \emph{Effective versus classical geometry:}  In the `interior region' the effective geometry is  \emph{very} different from the classical one because quantum geometry corrections to Einstein's equations become crucial. As we saw in section \ref{s4.B}, these corrections can be regarded as providing an effective stress-energy tensor that violates the strong energy condition in the Planck regime, leading to singularity resolution. But we also saw that in the low curvature region near horizons, these quantum corrections become negligible and Einstein's equations provide an excellent approximation to the effective equations (Fig. \ref{fig:2}).

What is the situation in the exterior region? Since the effective theory includes $\hbar$-dependent quantum corrections, effective metric never agrees completely with the classical Schwarzschild metric no matter how  far one recedes from the horizon (just as general relativistic corrections to Newtonian theory are never zero for any physical system). However, quantum corrections are again negligible  already at the horizon of macroscopic black holes and become even smaller rapidly in the asymptotic region. For concreteness, let us consider classical and effective solutions for $m = 10^{4} M_{\rm{Pl}}$ and examine the differences in the two space-time geometries in the asymptotic region I. Using \texttt{MATHEMATICA}, one finds the following illustrative numbers:\\
(i) As Eqs. (\ref{momentat}) and (\ref{eq:pct}) show the horizon radius of the effective solution is slightly larger than that in the classical theory. The relative difference is only $\sim 10^{-15}$ and the corrections fall off as $m^{-8/3}$. So, for a solar mass black hole, the relative difference in the horizon radius is $\sim 10^{-106}$!\\
(ii) In the classical theory the Ricci tensor is identically zero. In the effective theory,  at the horizon the square of the Ricci scalar of the effective metric is given by $R^{2}_{\rm eff} \approx 9.4 \times 10^{-25}$ {in Planck units} (for $m= 10^{4}$). It increases slightly as one recedes from the horizon and reaches a maximum of $(R^{2})_{\rm eff}^{\rm max} \approx 1.2 \times 10^{-24}$ at $T\approx 0.13$  and then decays rapidly to zero. (Recall that the horizon corresponds to $T=0$;  in terms of the radial coordinate $r$,  $R^{2}_{\rm eff}$ reaches its maximum at $\approx 1.14 r_{\rm hor}$.)  More generally, $(R^{2})_{\rm eff}^{\rm max} \sim 10^{-8} (\lp/m)^{4}$. These are measures of absolute smallness of  $(R^{2})_{\rm eff}^{\rm max}$. Since the classical Kretschmann scalar at the horizon is $K_{\rm cl}  \approx 7.5 \times 10^{-17}$,  the relative smallness of the departure from Einstein's equations is $R^{2}_{\rm eff}/K_{\rm cl} \approx 10^{-8}$ at the horizon.\\
(iii) One can also consider the square of the Ricci tensor of the effective metric. At the horizon it is $[R_{ab}R^{ab}]_{\rm eff} \approx 4.7 \times 10^{-25}$ and it decays rapidly in the asymptotic region (again for $m=10^{4} M_{\rm{Pl}}$).\\
Thus, for macroscopic black holes,  the Schwarzschild solution is an excellent approximation to the effective space-time metric throughout region I and the approximation rapidly improves as one moves to the asymptotic region.  Therefore, in this approach, large quantum corrections outside the horizons of macroscopic black holes envisaged in some approaches  (see, e.g. \cite{sg}) do not arise.\medskip

$\bullet$ \emph{Non-amplification of mass:} The approach most closely related to ours is the `generalized $\mu_{o}$-scheme' introduced and analyzed in detail  in \cite{cs}.  That investigation explored  the effective geometry only in the interior region depicted by  the central diamond ${\rm B} \cup {\rm W}$ in Fig. \ref{fig:4}.  A key feature of that effective geometry is that the radius $r_{\rm W}$ of the white hole type horizon is very large compared the radius $r_{\rm B}$ of the black hole type horizon in the diamond, growing as $r_{\rm W} \approx r_{\rm B} (r_{\rm B}/\lp)^{3}$.  So if we start with $r_{\rm B} = 3\, {\rm km}$ --the Schwarzschild radius of a solar mass black hole-- we have $r_{\rm W} \approx 10^{93}$\,Gpc! This effect was interpreted as quantum gravity induced mass inflation (see Eq. (\ref{inflation})). 
However, the physical origin of this mass inflation has remained unclear.

In our approach, by contrast, the ratio of the two radii is very close to $1$ for macroscopic black holes:
\be \f{r_{\rm W}}{r_{\rm B}} = 1+ \mathcal {O}\Big[\Big(\f{\Delta}{m^2}\Big)^{\f{1}{3}} \,\,\ln \Big(\f{m^2}{\Delta}\Big)\Big].\ee
If we again consider a solar mass black hole, we have $r_{\rm W} \approx (3 + \mathcal{O}(10^{-25})) {\rm km}$. Since we have a smooth effective geometry connecting the interior and asymptotic regions, which furthermore agrees with the Schwarzschild metric to an excellent degree of approximation, we can calculate the ADM mass of this solution. Since there is a time translation Killing field in the asymptotic region we can relate the radius of the  black hole type horizon  to the ADM mass in asymptotic region I and the white hole type horizon to the ADM mass of asymptotic region III of Fig. \ref{fig:4} (using Komar integrals discussed below.)  Each of these two ADM masses is \emph{extremely} well approximated by the radius of the corresponding horizon (divided by $2G$). Therefore we conclude that for macroscopic black holes the  ADM mass in all asymptotic regions are the same to an excellent approximation which, furthermore,  improves as the mass increases.\medskip

$\bullet$ \emph{Translational Killing vector and the Komar mass:}  By construction, the effective geometry admits a translational Killing field $X^{a}$ which, as in the classical theory, is time-like in exterior regions and space-like in the interior. Let us focus on the interior and calculate the Komar integrals 
\be \label{komar} K_{X}[S] := -\f{1}{8\pi G}\, \oint _{S}\epsilon_{ab}{}^{cd} \nabla_{c}X_{d} \, \d S^{ab} \ee
using various round 2-spheres $S$. Recall that, if  $S_{1}, S_{2}$ are joined by a 3-surface $M$, then
\be \label{balance} K_{X}[S_{2}] - K_{X}[S_{1}] = 2 \int_{M}  \big(\mathfrak{T}_{ab} - \f{1}{2}\mathfrak{T}  g_{ab}\big) X^{a} \d V^{b}, \ee
where $\d V^{b}$ is the oriented volume element of the 3-surface $M$. If we choose $S$ to lie on a horizon, the Komar mass $K_{X}[S]$ is related to the horizon radius by $2 G K_{X}[S] = r_{hor}$. Let $S_{1}$ lie on the black hole type horizon and $S_{2}$ lie on the  
white hole type horizon in the central diamond of Fig. \ref{fig:4} and $M$ be a 3-dimensional `tube' joining them.  Then we appear to have a paradox. On the one hand, in the interior region there is an effective stress-energy tensor because the quantum corrected metric is not Ricci flat. 
Therefore, the integrand on the right side in (\ref{balance}) is non-zero. Indeed,  the 3-surface $M$ must cross the transition surface $\T$ and, as we saw in section \ref{s4.B}, the energy density and pressures are quite large near the transition surface. Furthermore as Fig \ref{fig:3} shows, both these quantities are negative almost everywhere in the interior (except near the horizon where their positive values are quite small). Therefore, one would expect the integral on the right side of (\ref{balance})  to be negative and rather large. How could the two horizons then have the same mass  (to an excellent degree of approximation)?

The solution of this puzzle is conceptually interesting. The right hand side of (\ref{balance}) is indeed negative and large. But the effective geometry is such that it is given by $-2M_{\rm B}$, where $M_{\rm B} = r_{B}/2G$. Therefore while  $K_{X}[S_{1}] = M_{\rm B}$, we have $K_{X}[S_{2}] = - M_{\rm B}$ (to an excellent degree of approximation). Geometrically, the negative sign arises simply because  while the Killing vector $X^{a}$ is future directed along the black hole type horizon (and in the asymptotic region I), it is \emph{past directed} along the white hole type horizon (and in asymptotic region III)!  As is evident from Fig. \ref{fig:3}, this must happen simply because the effective metric and its Killing field $X^{a}$ are smooth. On the other hand, the ADM energy is defined at spatial infinity in each asymptotic region using the asymptotic Killing field which is \emph{future directed and unit} at spatial infinity and is thus positive. Thus, the effective solution has the striking property it introduces just the right type of effective stress-energy that, on the one hand, large enough to resolve the singularity and, on the other, achieves the fine balance that is needed to satisfy the following relations:
\be M_{\rm ADM}^{(I)} = M_{\rm ADM}^{(III)} \quad \hbox{\rm which requires} \quad K_{X}[S_{2}]
=- K_{X}[S_{1}].\ee   
Here $M_{\rm ADM}^{(I)}$ is the ADM mass  in the region I  and $M_{\rm ADM}^{(III)}$ the ADM mass in region III and  $K_{X}[S_{1}]$ is the Komar integral on the black hole type horizon and $K_{X}[S_{2}]$ on the white hole type horizon. It is thanks to this fine balance that there is no mass amplification in the large $m$ limit.

\section{Discussion}
\label{s6}

The issue of the fate of black hole singularities in quantum gravity has drawn a great deal of attention especially over the past two decades.  While there is  broad consensus that  singularities are {windows} onto physics beyond Einstein's theory, there is no general agreement on how the singularities are to be resolved and even whether one should expect them to be resolved. For example,  in the commonly used Penrose diagram of an evaporating black hole --first introduced by Hawking \cite{swh} over 40 years ago-- a singularity constitutes part of the future boundary of space-time even after the black hole has completely disappeared. Although this scenario is not based on a hard calculation in any approach to quantum gravity, it continues to be widely used. There is also a debate on whether quantum gravity effects associated with black holes would be important at horizons of macroscopic black holes and even in the exterior region well outside the horizons \cite{sg}.

In this paper we focused on a specific issue by restricting ourselves to the Kruskal space-time:
Is there a coherent, effective description that incorporates sufficient elements of a deeper quantum gravity theory to lead to the resolution of singularities of this space-time? If there is, further questions arise. What is the nature of the resulting quantum extension? Do the large quantum gravity effects that are needed to resolve the central singularity persist even in low curvature regions, thereby modifying the classical geometry near and outside the black hole horizon? Is the quantum corrected effective geometry well defined both in the `interior' region bounded by horizons as well as the `exterior' asymptotic region? If the extension includes anti-trapped regions, are they connected to new asymptotic regions? Is the ADM mass in these regions essentially the same as the initial mass one starts with or there is a significant mass inflation or deflation? We were able to answer these questions in detail within an effective theory that incorporates key elements of Riemannian quantum geometry  underlying LQG. 

The salient features of this effective description are the following. First, as shown in sections \ref{s2} and \ref{s3}, the black hole singularity is naturally resolved due to quantum geometry effects, i.e., because there is an area gap $\Delta$ in LQG. The singularity is replaced by a transition surface $\T$ which separates a trapped region to its past from the anti-trapped region to its future: Our effective description extends the Schwarzschild interior to include a white hole type horizon beyond the `would be' singularity. Since the effective metric is smooth, all curvature invariants are bounded. Furthermore --as is common in loop quantum cosmology-- each curvature invariant has an \emph{absolute} upper bound that does not grow with mass (section \ref{s4.C}). The expressions of these upper bounds contain \emph{inverse} powers of the area  gap $\Delta$. This is analogous to the observation --often emphasized by John Wheeler-- that  $\hbar$ appears in the denominator of the expression of the ground state energy of the hydrogen atom and the fact that it is non-zero prevents the minimum energy from being unbounded below.  Thus, there is a precise and sharp sense in which singularity resolution is due to quantum geometry effects that give rise to a non-zero area gap. While quantum corrections lead to large violations of Einstein's equations near the transition surface (section \ref{s4.B}) they become negligible in the low curvature region (section \ref{s4.D}). In particular, for macroscopic black holes with $M \gg M_{\rm Pl}$,  classical general relativity continues to provide an excellent approximation near and outside their horizons. While previous works focused only on the Schwarzschild interior, a key new feature of our analysis is that we were able to construct a Hamiltonian description and analyze effective dynamics also in the `exterior' region between the horizons and infinity (section \ref{s5.A}). These are joined in a smooth manner to the `interior' regions across horizons (section \ref{s5.B}), providing us with a quantum extension of the \emph{full} Kruskal space-time shown in Fig. \ref{fig:4}. 

There is a large body of work on the Kruskal interior within LQG \cite{ab,lm,bv,dc,cgp,bkd,cs,djs,oss,cctr}, most of which focuses on effective dynamics as in this paper. In all these investigations the black hole singularity is resolved. However,  as discussed in section \ref{s4.D},  there are also major differences from our approach. Physical results in \cite{ab,lm,cgp} can depend on fiducial structures that are introduced in the construction of the classical phase space, whence the details of their predictions have no invariant significance. Our approach also starts with fiducial structures to make various mathematical expressions well-defined.  However, all our final results are insensitive to these choices. The final results in approaches introduced in \cite{bv,dc,cs,oss} are also free of dependence on fiducial structures. However, they lead to large quantum effects in low curvature regions. For example,  for large black holes, the quintessentially  quantum transition surface $\T$ can emerge in regions with arbitrarily small curvature in some approaches \cite{cs}, while quantum dynamics drives the effective trajectories to regions of phase space where the basic underlying assumptions are violated in others \cite{bv,dc,cctr}. This does not occur in our approach. Indeed, this effective description is free from all known weaknesses of previous investigations of Kruskal space-time within LQG. 

Finally, another key difference from previous investigations is the following. They considered only the Schwarzschild `interior' and treated it as a homogeneous (Kantowski-Sachs)  cosmology, emphasizing issues that are central to anisotropic cosmological models. For example  some allowed matter \cite{dc2,js} and/or a cosmological constant  \cite{bkd,djs}, thereby taking the focus away from the Schwarzschild interior.  As mentioned already,  our effective theory encompasses both the interior and the asymptotic regions  and our focus is on black hole aspects such as trapped and anti-trapped surfaces in the `interior' region and properties of the ADM mass in the asymptotic region. A striking feature of this effective description is that, in the large mass limit, the ADM mass does not change as one evolves from one asymptotic region to another one to its future. This feature arises from a surprising aspect of the  specific way Einstein's equations receive quantum corrections. On the one hand, these corrections are large enough to create a sufficiently strong repulsive behavior that is needed to resolve the singularity. On the other hand, in the evolution from the black hole type horizon to the white hole type horizon, the violation of Einstein's equations is subtle: the effective stress-energy induced by quantum geometry just flips the sign of Komar mass, keeping its magnitude the same. This flip goes hand in hand with the change of orientation of the translational Killing field, which in turn assures that the ADM mass remains the same from one asymptotic region to another one to its future (section \ref{s5.B2}).   This is why  the geometry of the `interior' region  is symmetric under time reflection around the transition surface $\T$ (in the large mass limit).  A symmetric behavior has been sought after and achieved using phenomenological inputs before \cite{oss};  it has been postulated in studies on black hole to white hole transition \cite{hhcr}; and arrived at by imposing physically motivated conditions on the black hole evaporation time scale \cite{bcdhr}. In our approach it emerges from detailed effective dynamics and is more subtle. In particular,  \emph{exact} symmetry holds only in the infinite mass limit.

Our effective dynamics also provides a concrete context to compare and contrast expectations based on the quantum nature of Riemannian geometry a la LQG and those based on the  AdS/CFT type arguments.  Since the bulk/boundary duality proposed in the AdS/CFT correspondence has been verified in a large number of examples,  expected physical properties of quantum field theories on the boundary have been used to argue that quantum gravity will/should \emph{not} resolve certain bulk singularities, including those of  the classical  Schwarzschild-Anti-de Sitter space-times \cite{eh}.   Note that these conclusions on the nature of bulk geometry  are indirect in that they are arrived at starting from physically desirable properties of theories on the boundary, assuming the boundary/bulk correspondence.  By contrast, in LQG one works directly with the bulk. Since our effective theory does resolve  Schwarzschild singularities in a coherent fashion, there is tension between the two sets of ideas.  There is no  contradiction since the plausibility arguments of \cite{eh} make a strong use of asymptotically Anti-de Sitter boundary conditions and do not apply to the asymptotically flat situation we have considered. Therefore, it would be of interest to see if the effective theory proposed here can be extended to the asymptotically Anti-de Sitter case. A result in either direction will provide valuable guidance.

We will conclude by pointing out some important limitations of our analysis.  As in the previous investigations, it is straightforward to introduce the kinematical Hilbert space of states by exploiting the underlying homogeneity. Furthermore, using considerations of Appendix \ref{a1}, one can write down the quantum Hamiltonian constraint. However, its explicit action is rather complicated. The situation was initially the same with the `improved dynamics' scheme in LQC, where it took some effort  \cite{aps3}  to simplify the action of the Hamiltonian constraint sufficiently to make subsequent calculations manageable. The simplified form could  then be used to arrive systematically at the quantum corrected, effective equations \cite{jw,vt,asrev}.  For the Kruskal black holes now under consideration, one would similarly have to first simplify the action of the Hamiltonian constraint significantly to `derive' the effective equations proposed in this paper starting from the quantum theory. Secondly, the question of stability of our effective space-times has not been investigated. This is a difficult issue because we do not have  quantum corrected equations for full general relativity. Nonetheless, since significant progress has been made on cosmological perturbations propagating on quantum FLRW geometries \cite{aa-ijmpd,apbook}, it may be possible to analyze this issue  in detail. A more important limitation comes from the fact that our analysis is confined to the \emph{eternal} black-white holes of Kruskal space-time. To address key conceptual issues such as the possibility of information loss, one would  have to consider black holes formed by gravitational collapse. For these situations, as in classical general relativity, only a small part of  the Penrose diagram of Fig. \ref{fig:4} will be relevant. One would have dynamical horizons which are either space-like (in the classical phase when the black hole grows) or time-like (during the quantum evaporation process), rather than null as in the Kruskal picture considered here; only a finite portion of the transition surface will appear because of the black hole evaporation; and there will likely be only one asymptotic region (see, e.g. \cite{aamb}). Thus,  the conceptual structure of the framework would be very different  from the full extension of Kruskal  space-time introduced here. Nonetheless, portion of this extended space-time will be relevant to the analysis and may in fact suffice to address deep conceptual puzzles that arise already in the semi-classical regime, e.g., during the phase in which a solar mass black hole shrinks to lunar mass due to evaporation \cite{aa-ilqg,cdl}.  Furthermore,  just as the analysis of quantum fields on Kruskal space-time provided useful techniques in the analysis of the Hawking process for physically more realistic collapsing situations, techniques developed in this quantum extension of Kruskal space-time should be helpful for the much more interesting physical problem of the fate of black hole singularities in dynamical processes.

\section*{Comments on a recent misleading Note}
 
A recent note \cite{bmm1} contains some misleading remarks on the work reported in this paper. The purpose of this short addendum is to remove the possible confusion it may lead to.

From the main body of the paper,  our goals are clear. Our discussion is limited to the specific case of a Schwarzschild black hole and we seek quantum parameters $\delta_{b}, \delta_{c}$ such that:\\ 
(i) They are phase space functions that remain constant along effective dynamical trajectories; \\
(ii) these trajectories are long enough so as to contain the transition surface $\T$ that separates the trapped region from the anti-trapped region in the effective geometry of  Schwarzschild interior;\\
(iii) on the transition surface $\T$, the quantum area conditions (\ref{area2}) and (\ref{area3}) are satisfied. \vskip0.1cm
In Appendix A we develop a framework to achieve these highly non-trivial goals. The framework requires one to execute three steps that have been spelled out in the paragraph after Eq.  (\ref{H2}). The third of these steps requires gauge fixing. As with any system with first class constraints, gauge fixing conditions have to be globally well-defined so that the dynamical trajectory does not stop because of a bad choice. In addition, in accordance with our original goal, the resulting dynamical trajectories have to satisfy conditions (i) - (iii),  spelled out above. In Appendix \ref{a1} we show in detail how the three steps can be executed so that the resulting effective trajectories  meet conditions (i) -(iii). Specifically, constraints of the form (\ref{A9}) ensure (i), and the more specific choice (\ref{db-dc2}) of these constraints together with the gauge fixing conditions (\ref{A16}) ensure (ii) and (iii) for macroscopic black holes considered in the paper.  It is clear from this discussion that one would completely miss the boat if one ignores the main goals (i) - (iii)  while carrying out the technical steps of Appendix \ref{a1}.

But this is precisely what is done in sections 2.2 and 3 of \cite{bmm1}. First, dynamical trajectories in these sections end abruptly because gauge fixing goes bad. More importantly, there is no analysis of whether the key condition (iii) above is satisfied! It is quite delicate to meet this requirement and general considerations suggest that it is highly unlikely that they are satisfied on the dynamical trajectories discussed in section 3 of \cite{bmm1}.  Therefore these arguments  do not shed any useful light on our analysis. In particular, the statements in \cite{bmm1} about `errors' are based on gauge choices that fail to meet conditions (i) - (iii) that are central to our analysis.

We do not claim that the choices made in the Appendices of our manuscript are the only ones that lead to dynamical trajectories meeting our goals.  It is often the case in physics that initially one has to make certain choices to execute a program. If they lead to physically interesting results, one gains confidence which in turn motivates further work that can lead to uniqueness results, that then justify the initial choices. This has happened repeatedly in LQG as well as other areas of physics. For example, the representations of the holonomy flux algebra that are used both in LQC and LQG were first introduced by  first choosing a certain measure on the quantum configuration space \cite{al2,abl}. Uniqueness results justifying this choice followed several years later \cite{lost,cf,aamc,eht}. Indeed, even the Fock representation for free quantum fields in Minkowski space-time had been in use for a couple of decades before Segal \cite{segal} and others established the uniqueness result.
Similarly, the $\bar\mu$-dynamics in homogeneous isotropic LQC was introduced by using well-motivated choices \cite{aps2} but initially it was far from being clear that there aren't other viable choices as well. It took several years to establish that the $\bar\mu$ scheme is singled out uniquely by certain physical criteria \cite{uniq1,uniq2}. In the Schwarzschild case now under consideration, similar uniqueness results may follow for the choices we made. Or, it  may well be that there are other choices that also satisfy conditions (i) - (iii), we started out with. If so, it would be very interesting to compare and contrast them with our scheme. But we fail to see the point of floating possibilities that do not meet these requirements.
 
Finally, one of the several strengths of the approach presented in this paper is that, because our basic variables are connections, it is meaningful to consider holonomies which, in turn, systematically lead to the replacement of connection components  by their trigonometric functions. In other approaches, one sometimes simply converts phase space variables that are \emph{not} connection components to their trigonometric functions, and then passes over to a quantum theory (see, e.g., \cite{bmm2}). This is done  without any obvious justification or plausible contact with LQG. It would be an order of magnitude harder to justify these choices or prove uniqueness theorems for them.

\section*{Acknowledgments}  This work was supported in part by the  NSF grants PHY-1454832, PHY-1505411 and PHY-1806356,\, grant UN2017-9945 from the Urania Stott Fund of the Pittsburgh Foundation, the Eberly research funds of Penn State, and by Project. No. MINECO FIS2014-54800-C2-2-P from Spain and its continuation Project. No. MINECO FIS2017-86497-C2-2-P.  We thank Eugenio Bianchi and Hal Haggard for useful discussions.  We also acknowledge valuable correspondence with Norbert Bodendorfer that led to improvements in our presentation.

\begin{appendix}
\section{Quantum parameters as judiciously chosen Dirac observables} \label{a1}
Given the quantum parameters $\delta_{b}, \delta_{c}$,  the Hamiltonian constraint  of the effective theory is given by (see (\ref{H_eff})) : 
\begin{align} \label{H1} 
H_{\mathrm{eff}} =  - \f{1}{2 G \gamma} \Big[ \Big(\f{\sin (\delta_b b)}{\delta_b} + \frac{\gamma^2 \delta_b}{\sin(\delta_b b)} \Big) \, p_b + 2 \f{\sin (\delta_c c)}{\delta_c}  \, p_c  \Big]  \equiv \f{ L_{0}}{G}\,\, \Big[O_{1} - O_{2}\big].
\end{align}
\be {\rm where}\quad O_{1}:= - \f{1}{2\gamma}\, \Big[\f{\sin \delta_{b} b}{\delta_{b}} + \f{\gamma^{2} \delta_{b}}{\sin\delta_{b}b}\Big]\, \f{p_{b}}{L_{o}}, \quad {\rm and}\quad
O_2 := \Big[ \frac{ \sin\dc c}{\gamma L_{o}\dc}\Big]\, p_{c}. \label{H2}\ee
The task of making  $\delta_{b}, \delta_{c}$ constants of motion is technically subtle:  $\delta_{b}, \delta_{c}$ themselves feature in the expression (\ref{H1}) of the Hamiltonian constraint that determines the dynamical trajectories along which $\delta_{b}, \delta_{c}$ are to be constants. Thus we have to choose $\delta_{b},\delta_{c}$ astutely to ensure this internal consistency.  The goal of this Appendix is to show that these goals can be achieved and that the choice (\ref{db-dc}) made in section \ref{s4.A} satisfies this subtle consistency.\medskip


To achieve this goal,  we will proceed in the following steps:\\ 
(1) We will first extend the 4-dimensional phase space $\Gamma$ of the main text (with canonically conjugate variables $b,p_{b}$, and $c,p_{c}$) to a 8-dimensional phase space $\GammaE$ by introducing  2 additional \emph{independent} canonically conjugate pairs $\delta_{b}, p_{\delta{b}}$ and $\delta_{c}, p_{\delta{c}}$. Thus, in particular, on $\GammaE$, the \emph{`would be'} quantum parameters $\delta_{b}$ and $\delta_{c}$ are not functions of $b,p_{b}; c,p_{c}$, but Poisson commute with all original phase space variables.\\
(2)  We will consider the natural lift  $\Hee$ of the Hamiltonian $H_{\mathrm{eff}}$ to $\GammaE$ and examine the Hamiltonian flow it generates. As we argue below,  $O_{1}, O_{2}$ are Dirac observables for this flow. If $\delta_{b}$ is required to be a function only of $O_{1}$ and $\delta_{c}$ of only $O_{2}$ then $\delta_{b}, \delta_{c}$ would also be Dirac observables.
Our goal then is to introduce these dependences as two \emph{new constraints} such that they, together with the Hamiltonian constraint function $\Hee$ form a \emph{first class set}
on the extended phase space $\GammaE$.  Then, in particular, the Hamiltonian flow generated by  $\Hee$ on the extended phase space $\GammaE$ will be tangential to the 5 dimensional constraint surface $\GammaEb$.\\
(3) Finally, our goal is to choose two gauge fixing conditions \emph{for the newly introduced constraints} such that the 4-dimensional reduced phase $\GammaEh$  corresponding to these constraints is symplectomorphic to the original 4-dimensional phase space $\Gamma$ we began with. The dynamical flow on $\GammaEb$ would then be induced by the Hamiltonian $\Hee$, but with $\delta_{b},\delta_{c}$ given by the specified functions of $O_{1}$ and $O_{2}$ respectively. Assuming all requirements on the choice of new constraints and their gauge fixing can be satisfied, the dynamical flow on $\GammaEb$ will naturally descend to the constraint surface $\bar\Gamma$ of the original phase space $\Gamma$, providing us with the desired dynamics.
\\
Conditions in the second and third step are quite demanding and a priori it is not clear that they can be met. However, as we show below, there is a large class of choices of $\delta_{b}, \delta_{c}$ for step (2) for they can be made. Among them is the choice (\ref{db-dc}) made in section \ref{s4.A}. We will now carry out these three steps systematically.\medskip

The extended phase space $\GammaE$ is naturally coordinatized by 4 canonically conjugate pairs $b,p_{b};\, c, p_{c}; \, \delta_{b}, p_{\delta{b}}; \, \delta_{c}, p_{\delta{c}}$. Note that $\delta_{b},\delta_{c}$ are just new, independent canonical variables that Poisson commute with the original $b,p_{b}; c, p_{c}$ and their conjugate momenta $p_{\delta_{b}}, p_{\delta_{c}}$ do not appear in the expression of 
\be \label{Hee} \Hee = \f{L_{0}}{G}\,\, \Big[O_{1} - O_{2}\big].\ee
Since furthermore, $O_{1}$ refers only to the $b$-sector and $O_{2}$ only to the $c$-sector, it follows that $O_{1}, \, O_{2}$  Poisson commute and are, furthermore, Dirac observables of the flow generated by $\Hee$ on $\GammaE$. (As one would expect from our discussion in the main text, along dynamical trajectories  $O_{2}$ equals $O_{1}$ and they will turn out to be the mass $m$. See Eq (\ref{m}).) To carry out  steps (2) and (3) explicitly, it is convenient to first make a detour and introduce a canonical transformation so that $O_{1}, O_{2}, \delta_{b}, \delta_{c}$ are the new configuration variables, and their momenta are given by:
\begin{align}
P_1&=-\frac{2L_o}{Gb_o}\tanh^{-1}\left[b_o^{-1}\cos(\delta_b b)\right]- \frac{2L_o}{G}\ln\left|\frac{\gamma\delta_b}{2}\right|, \label{ct1}\\
P_2&= -\frac{L_o}{2G}\ln\left[\left|\frac{2p_{c}}{L_o\delta_c}\Big[ \frac{ \sin\dc c}{\gamma L_{o}\dc}\Big]\,\tan\left(\frac{\delta_c c}{2}\right)\right|\right],\label{ct2} \\\nonumber
P_{\delta_b} &= p_{\delta_b}-\frac{p_b}{2\gamma G}\left\{\frac{2}{\delta_b} \left[b -\left(\frac{\sin(\delta_b b)}{\delta_b}+\frac{\gamma^2\delta_b}{\sin(\delta_b b)}\right)\right]\right.\\
&\left.+\frac{\gamma}{b_o^2\sin(\delta_b b)}\left(2\gamma \cos(\delta_b b)+\frac{\gamma}{b_o}\tanh^{-1}\left[b_o^{-1}\cos(\delta_b b)\right]\left(2b_o^2-1-\cos(2\delta_b b)\right)\right)\right\}, \label{ct3}\\
P_{\delta_c} &= p_{\delta_c}-\frac{p_c}{2\gamma G \delta_c}\left(c-\frac{\sin(\delta_c c)}{\delta_c}\right). \label{ct4}
\end{align}
Direct calculations show that  Eqs (\ref{ct1}) - (\ref{ct4}) define a canonical transformation on $\GammaE$ from the original variables $(b,p_{b};\, c,p_{c};\,\delta_{b},p_{\delta_{b}};\, \delta_{c},p_{\delta_{c}})$\,\, to\,\, $(O_{1},P_{1};\, O_{2},P_{2};\, \delta_{b},P_{\delta_{b}};\, \delta_{c},P_{\delta_{c}})$:
\be \{O_i,P_j\}=\delta_{ij},\quad \{\delta_b,P_{\delta_b}\}=1,\quad \{\delta_c,P_{\delta_c}\}=1,\ee
and all remaining Poisson brackets vanish. This transformation is complicated  in part because we have also ensured that it is well-defined in the classical limit in which 
$\delta_b\to 0$ and $\delta_c\to 0$.  This transformation is invertible and we provide the explicit inverse at the end of this Appendix. 

The step (2) asks us to make $\delta_{b}, \delta_{c}$ the desired functions of other phase space variables by introducing constraints on $\GammaE$.  Since in the final picture we would like $\delta_{b}, \delta_{c}$ to be Dirac observables and since we know that $O_{1}$ and $O_{2}$ are Dirac observables, we choose these constraints to be: 
\be \Phi_1=O_1-F_b(\delta_b) \approx 0,  \quad {\rm and} \quad \Phi_2=O_2-F_c(\delta_c) \approx 0, \label{A9}\ee
where $F_{b}$ and $F_{c}$ are functions of $\delta_{b}, \delta_{c}$ whose functional can be quite general, subject to suitable regularity conditions.  (Our final choice (\ref{db-dc}) of these quantum parameters is of this form because, as noted above $O_{1} = m = O_{2}$ on dynamical trajectories.) Since $O_{1}, O_{2}$ do not depend on the momenta $P_{\delta_{b}}, \, P_{\delta_{c}}$ of the quantum parameters $\delta_{b}, \delta_{c}$, it follows immediately that the three constraints $\Hee \approx 0,\, \Phi_{1} \approx 0, \, \Phi_{2} \approx 0$ on $\GammaE$ constitute a first class system. Thus we have met the conditions specified in step (2).

Next, consider the flow of the \emph{total} Hamiltonian:
\ba \label{HT} H_{\rm T}^{\rm ext}  &=& \underbar{N} \Hee + \lambda_{1} \Phi_{1} + \lambda_{2} \Phi_{2}\nonumber\\ 
&=& - \f{L_o\underbar{N}}{G} \left[O_2-O_1\right] +\lambda_1 \left[O_1-F_b(\delta_b)\right]+\lambda_2\left[O_2-F_c(\delta_c)\right], \ea
where $\underbar{N}, \lambda_{1}, \lambda_{2}$ are Lagrange multipliers. The equations of motion defined by the Hamiltonian flow of  $H_{T}^{\rm ext}$ are
\begin{align}
\dot O_1 &= 0,\quad \dot P_1 = -\f{L_o\underbar{N}}{G}-\lambda_1;\qquad
\dot O_2 = 0,\quad \dot P_2 = \f{L_o\underbar{N}}{G}-\lambda_2; \label{evo1}\\
\dot \delta_b &= 0,\quad \dot P_{\delta_b}=\lambda_1F_b'(\delta_b); \qquad\quad\,\,
\dot \delta_c = 0,\quad \dot P_{\delta_c}=\lambda_2F_c'(\delta_c); \label{evo2}
\end{align}
and, in addition, the phase space variables are subject to the three constraints:
\begin{align}
O_1-O_2\,\,\approx\,\,0,\quad O_1-F_b(\delta_b)\,\,\approx\,\,0, \quad O_2-F_c(\delta_c) \,\,\approx\,\, 0.
\end{align}
The equations are thus very simple; this is the reason why we introduced the new canonical variables.  We  know from general arguments that the flow is tangential to the 5-dimensional constraint surface $\GammaEb$ in $\GammaE$. Equations of motion make this explicit. \\  

\emph{Remark:} The explicit form of solutions implies that  $\delta_{b}, \delta_{c}$ are constants along dynamical trajectories, as desired. Furthermore, for any choice of (regular) functions $F_{b}, F_{c}$, their explicit dependence on the new configuration variables $O_{1}$ is known. The construction of section \ref{s4.A} led us to set 
\be\label{db-dc2}                                                    
 F_b(\delta_b)=\Big(\frac{\sqrt{\Delta}}{\sqrt{2\pi}\gamma^2\delta_b^3}\Big), \qquad          
F_c(\delta_c)=\frac{1}{8} \Big(\frac{\gamma\Delta^2}{4\pi^2 (L_{o}\delta_c)^3}\Big),         
\ee
but the main conclusions of this Appendix do not require this specific choice. Discussion of the causal structure of of the Kruskal interior of section \ref{s3} holds for the class of quantum parameters that correspond to general $F_{b}$ and $F_{c}$ --and these include the choices made in the `generalized $\mu_{o}$ approaches' \cite{cs,oss}, discussed in section \ref{s4.D.3}.
\\
 
But we still need to relate the dynamical trajectories on $\GammaEb$ to those on the constraint surface $\bar\Gamma$ of the original phase space $\Gamma$. This requires completion of step (3) of the program: Introduction of gauge fixing for the new constraints $\Phi_{1} \approx 0$ and $\Phi_{2} \approx 0$ so that the resulting 4-dimensional reduced phase space $\GammaEh$ is symplectomorphic to the original phase space $\Gamma$ spanned by $b,p_{b}\, c, p_{c}$. This means that the gauge fixing conditions should be such that the terms
\be \left({\rm d}\delta_b\wedge {\rm d}P_{\delta_b}+{\rm d}\delta_c\wedge {\rm d}P_{\delta_c}\right) \ee
in the expression of the symplectic structure on $\GammaE$ should vanish when pulled back to $\GammaEh$. An examination of the form of the constraints leads us to conditions of the form
\begin{equation}\label{A16}
P_{\delta_b}=G_b(O_1,O_2), \,\,\, P_{\delta_c}=G_c(O_1,O_2),\quad \hbox{{\rm such that}}\quad
\frac{1}{F_b'(\delta_b)}\frac{\partial G_b}{\partial O_2}=\frac{1}{F_c'(\delta_c)}\frac{\partial G_c}{\partial O_1}.
\end{equation}
The form of the evolution equations implies that these are good gauge fixing conditions in the sense that each gauge orbit generated by the new constraint functions $\Phi_{1}$ and $\Phi_{2}$ intersects the gauge fixed surface once and only once. Finally  evolution consistent with this gauge fixing is obtained by setting $\lambda_{1} =\lambda_{2} =0$ in the expression (\ref{HT}) of  $H_{\rm T}^{\rm ext}$.  Thus, we have exhibited a family of gauge conditions satisfying conditions (3) in our prescription.

On the 4-dimensional reduced phase space  $\GammaEh$, then, dynamics is generated by the Hamiltonian constraint $\Hee$ of (\ref{Hee}) where, now, $\delta_{b}, \delta_{c}$ are determined by the constraints $\Phi_{1} =0$ and $\Phi_{2} =0$. The explicit form of this evolution is given by the restriction of (\ref{evo1}) and (\ref{evo2}) to the constraint surface $\Hee =0$ of the reduced phase space $\GammaEh$.  Evolution of $P_{1}, P_{2}$ is extremely simple, and $O_{1}, O_{2};\,\, \delta_{b}, \delta_{c}$ are constants of motion related via constraints $\Phi_{1} =0, \Phi_{2}=0$.

Finally, one can rewrite these evolution equations in terms of $b,p_{b};\, c, p_{c}$ by using the inverse of the canonical transformations (\ref{ct1}) - (\ref{ct4}) on $\GammaE$:
\begin{align}
\cos(\delta_b b) &=-b_o \tanh\left[\frac{b_o}{2}\left(\frac{G P_1}{L_o}+2\ln\left|\frac{\gamma \delta_b}{2}\right|\right)\right],\\  
\frac{b_{o}^{2}}{2L_{o}\gamma} {p_b} &= - \delta_b O_1{b_o^2}\,\ {\cosh\left[\frac{b_o}{2}\left(\frac{G P_1}{L_o}+2\ln\left|\frac{\gamma \delta_b}{2}\right|\right)\right]^2 \sqrt{1 - b_o^2 \tanh\left[\frac{b_o}{2}\left(\frac{G P_1}{L_o}+2\ln\left|\frac{\gamma \delta_b}{2}\right|\right)\right]^2}},\\
\tan\left(\frac{\delta_c c}{2}\right) &= \pm \frac{L_o\delta_c}{2O_2}e^{-2 G P_2/L_o},\\
p_c &= \gamma O_2^2 \left(e^{2 G P_2/L_o}+\frac{L_o^2\delta_c^2}{4O_2^2}e^{-2 G P_2/L_o}\right).
\end{align}
 (In the effective theory the phase space $\GammaE$  is restricted such that  $c\in[-\pi/\delta_c,0) \cup(0,\pi/\delta_c]$ and $b\in[-\pi/\delta_b,0)\cup(0,\pi/\delta_b]$ (see footnote 
 \ref{fn6})).
The resulting equations of motion for $b,p_{b}; c, p_{c}$ are precisely Eqs. (\ref{eom1}) and (\ref{eom2}) of the main text. \\

\emph{Remark:}  The Hamiltonian constraint  (\ref{H1}) can be formally promoted to an operator $\hat{H}$ on the Kinematical Hilbert space $\H_{\rm kin}$ used in the literature (see, e.g. \cite{ab,cs}), but its explicit action on the basis states normally used is rather complicated because $\delta_{b}, \delta_{c}$ depend on the phase space variables via Eq. (\ref{db-dc}).  A possible avenue to simplify the action is suggested by the strategy adopted in this Appendix. One may extend the kinematical Hilbert space by introducing new degrees of freedom corresponding to $\delta_{b}, \delta_{c}$ (and their canonically conjugate variables) also in the quantum theory, and then impose the three first class constraints as operator equations on the extended kinematical Hilbert space $\H_{\rm kin}^{\rm ext}$. This step would be straightforward, e.g., if one uses the representation in which $p_{b}, p_{c}, \delta_{b}, \delta_{c}$ are diagonal. However, further work is need to first explicitly solve the new operator constraints $\hat\Phi_{i} =0$  since $\delta_{b}, \delta_{c}$ also appear in the expressions of $\hat{O}_{i}$ ($i=1,2$). One possibility is to first seek the generator of the canonical transformations (\ref{ct1})  - (\ref{ct2}), promote it to an operator to go back and forth between representations in which $p_{b}, p_{c}, \delta_{b}, \delta_{c}$ are diagonal and the one in which 
$O_{1}, O_{2}, \delta_{b}, \delta_{c}$  (or, $P_{1}, P_{2}, \delta_{b}, \delta_{c}$)  are diagonal and exploit the simplicity of constraints in terms of $O_{1}, O_{2}, \delta_{b}, \delta_{c}$.

\section{Determining the quantum parameters $\delta_b$ and $\delta_c$}
\label{a2}

Results of Appendix \ref{a1} hold for a judiciously chosen but still a large class of the quantum parameters $\delta_{b}, \delta_{c}$. In the main body of the paper, we used a specific form \eqref{db-dc}.  In this Appendix we derive this equation starting from conditions \eqref{area3} and \eqref{area4} on the area enclosed by the chosen plaquettes on the transition surface $\T$. We will first derive analytical expressions in the large $m$ limit  and then discuss some subtleties associated with the exact solutions. The strategy is to first obtain expressions of $p_{b}|_{\T}$ and $ p_{c}|_{\T}$ at the transition surface $\T$ as functions of $\db,\dc$ using explicit solutions \eqref{eq:pb} and \eqref{eq:pc} to effective equations, and then determine the two unknowns $\db,\dc$ using the two area conditions \eqref{area3} and \eqref{area4}.

The expression of $p_{c}$ at the transition surface is simple: $p_c|_{\cal T}=m(\gamma L_o\delta_c)$. By contrast, the expression of $p_b|_{\cal T}$ is intricate and far more non-trivial.  To gain control over this expression, let us first consider an expansion in the limit $\db\ll1$ and $L_o\dc\ll \sqrt{\Delta}$. The leading order gives  $p_b|_{\cal T} \simeq L_o(2 m^3 L_o \gamma \dc)^{1/4}$. Then, we can solve Eqs. \eqref{area3} and \eqref{area4}, and obtain
\be
\db\propto \Big(\f{\sqrt{\Delta}}{m}\Big)^{1/3} \quad {\rm and} \quad \frac{L_o\dc}{\sqrt{\Delta}}\propto 
\Big(\f{\sqrt{\Delta}}{m}\Big)^{1/3}.  
\ee
(Recall that it is the combination $L_{o} \dc$ that is invariant under the rescaling $L_{o}  \to \alpha L_{o}$ of the fiducial cell.) These conditions imply that $\db$ and $L_o\dc$ are Dirac observables: they are constant on a given solution, but they can change from one solution to another.  Our initial assumption $\db\ll 1$ and $L_o\dc\ll \sqrt{\Delta}$ is automatically satisfied in the desired large $m$ limit,  $m\gg m_{\rm Pl}$,  if  we set
\be\label{eq:alp-bet}
\db=A\,\, \Big(\frac{\sqrt{\Delta}}{m}\Big)^{1/3} \quad  {\rm and} \quad \f{L_o\dc}{\sqrt{\Delta}} =  B\,\, \Big(\frac{\sqrt{\Delta}} {m}\Big)^{1/3},
\ee
where $A$ and $B$ are some dimensionless constants (independent of the mass $m$) which get  determined from the minimum area conditions.

Using Eq. \eqref{area4} we find that $A$ and $B$ are related via
\be\label{eq:beta}
B = \frac{1}{4\pi \gamma A^2},
\ee
enabling us to trade $B$ for $A$.  Finally, to determine $A$ we use the expression of $p_b|_{\cal T}$ and (\ref{eq:beta}).  A straightforward computation  provides us with an equation for $A$ in the large $m$ limit:
\be
\frac{A^{3/2} \gamma}{2^{1/4}(4\pi)^{3/4}}+\frac{2^{1/4}}{(4\pi)^{5/4}A^{3/2} \gamma} = \frac{1}{2\pi}.
\ee
This equation has only one real solution given by
\be\label{eq:alphasol}
A=\Big(\frac{1}{\sqrt{2\pi}\gamma^2}\Big)^{1/3} 
\ee
which then determines  $B$ via (\ref{eq:beta})
\be\label{eq:betasol}
B=\frac{1}{2}\Big(\frac{\gamma}{4\pi^2}\Big)^{1/3}.
\ee
One can easily see that these values of $A$ and $B$, together with \eqref{eq:alp-bet}  provide  the expression \eqref{db-dc} of $\db,\dc$
\be
\db=\Big(\frac{\sqrt{\Delta}}{\sqrt{2\pi}\gamma^2m}\Big)^{1/3}, \qquad 
L_{o}\dc=\frac{1}{2} \Big(\frac{\gamma\Delta^2}{4\pi^2 m}\Big)^{1/3}.
\ee
given in the main text.  Thus, in the  $m\gg \lp$ limit, solutions to the area conditions  \eqref{area3} and \eqref{area4} are given by \eqref{db-dc} to leading order.
\nopagebreak[3]\begin{figure} [h] 
\includegraphics[width=0.8\textwidth]{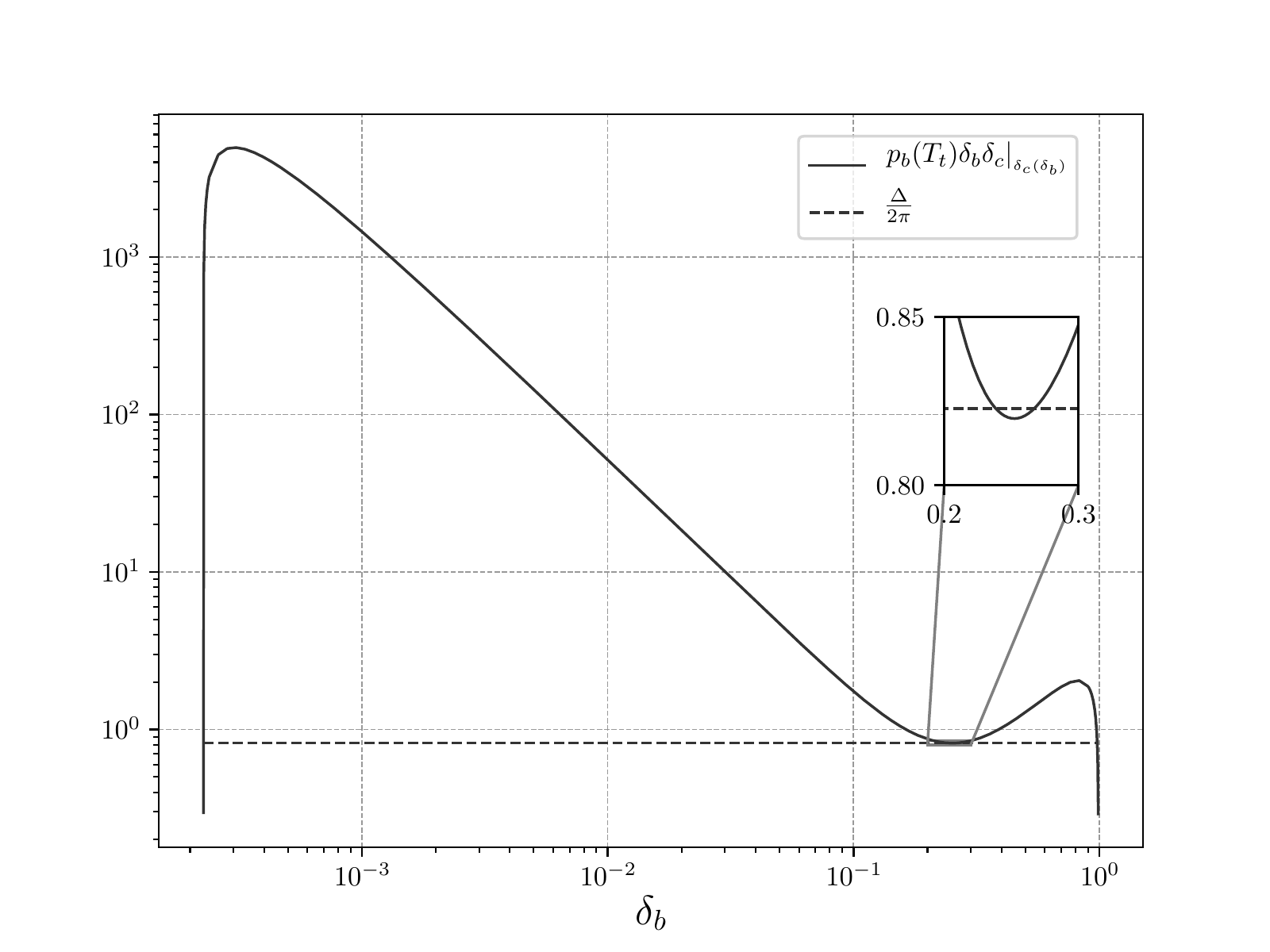}  
\caption{\footnotesize{The roots of $\delta_b$ obtained from solving Eqs. \eqref{area3} and \eqref{area4}. In the large $m$ limit, the central roots are extremely well approximated by Eqs. \eqref{eq:alphasol} and \eqref{eq:betasol}. The leftmost and rightmost roots turn out to have unphysical properties.  } }
\label{fig:5}
\end{figure}
Finally, it is instructive to solve Eq. \eqref{area4} numerically,  \emph{without taking the large $m$ limit}. Since we know $p_c|_{\cal T}=m(\gamma L_o\delta_c)$,  let us start by first solving Eq. \eqref{area4} for $L_o\dc$ and use the solution  in Eq. \eqref{area3}. Then the only unknown in the solution (\ref{eq:pb}) for $p_{b}$ is $\db$. Therefore we can numerically evaluate the left side $p_{b}(T_{\T} )\db\dc$ of (\ref{area3}) as a function of $\db$. The solid curve in  Fig \ref{fig:5} plots this function for $m =10^4$. The right side of (\ref{area3})  is a constant,  $\Delta/2\pi$, shown by the dashed line. The two curves have 4 intersections that represent 4 roots of our equation for $\db$. (Thus, if we do not take the large $m$ limit, the two conditions (\ref{area3}) and (\ref{area4}) do not  quite determine the unknowns $\db,\dc$ uniquely; we are left with a discrete, 4-parameter family of degeneracy.) We will refer to the 4 roots as as the leftmost, the two central and the rightmost.  Their properties can be summarized as follows. The two central roots are the relevant ones for our analysis. In the large mass limit,  they approach each other and rapidly converge to a single degenerate value, given by the analytic expression \eqref{db-dc}.  This root corresponds to the large $m$ values of constants $A$ and $B$ given in \eqref{eq:alphasol} and \eqref{eq:betasol}. For macroscopic black holes it is these central roots that yield the effective geometries discussed in sections \ref{s4.B} and \ref{s4.C}.

 The leftmost and the rightmost roots, on the other hand, are unphysical. The leftmost root gives a value for $\db$ that decreases rapidly as a function of the mass (faster than $m^{-1/3}$) while $\dc$ grows monotonically. In this case, the effective dynamics results in large quantum corrections at the black hole type horizon. For the rightmost root,  both $\db$ and $\dc$ decrease with the mass, however $\db$ does it very slowly. Although quantum corrections are small close to the black hole type horizon, they grow very quickly and become important while the Kretschmann scalar is still small. Thus, the leftmost and the rightmost roots can not yield physically viable solutions. That is why we focused on the limiting value of the central roots in our analysis in the main text. Finally, for macroscopic black holes, the two central roots are extremely close to one another, whence corrections to the asymptotic value are negligible. 
\goodbreak
\end{appendix}

\end{document}